\documentstyle[12pt,epsf,amssymb]{article} 
\catcode`@=11
\newcount\@tempcntc
\def\@citex[#1]#2{\if@filesw\immediate\write\@auxout{\string\citation{#2}}\fi
  \@tempcnta\z@\@tempcntb\m@ne\def\@citea{}\@cite{\@for\@citeb:=#2\do
    {\@ifundefined
       {b@\@citeb}{\@citeo\@tempcntb\m@ne\@citea\def\@citea{,}{\bf ?}\@warning
       {Citation `\@citeb' on page \thepage \space undefined}}%
    {\setbox\z@\hbox{\global\@tempcntc0\csname b@\@citeb\endcsname\relax}%
     \ifnum\@tempcntc=\z@ \@citeo\@tempcntb\m@ne
       \@citea\def\@citea{,}\hbox{\csname b@\@citeb\endcsname}%
     \else
      \advance\@tempcntb\@ne
      \ifnum\@tempcntb=\@tempcntc
      \else\advance\@tempcntb\m@ne\@citeo
      \@tempcnta\@tempcntc\@tempcntb\@tempcntc\fi\fi}}\@citeo}{#1}}
\def\@citeo{\ifnum\@tempcnta>\@tempcntb\else\@citea\def\@citea{,}%
  \ifnum\@tempcnta=\@tempcntb\the\@tempcnta\else
   {\advance\@tempcnta\@ne\ifnum\@tempcnta=\@tempcntb \else \def\@citea{--}\fi
    \advance\@tempcnta\m@ne\the\@tempcnta\@citea\the\@tempcntb}\fi\fi}
\catcode`@=12
\begin{document}
\vskip -1.5cm

\begin{center}

{\large 
{\bf Realizations of Hybrid Inflation in Supergravity }}\\[0.3cm]
{\large {\bf  with natural initial conditions}}\\[1.4cm]
{\large C. Panagiotakopoulos}\\[0.4cm]
{\em Physics Division, School of Technology,
         Aristotle University of Thessaloniki,\\
         54124 Thessaloniki, Greece}
\centerline{and}
{\em Department of Physics, King's College London, Strand, London WC2R 2LS, UK}
\centerline{and}
{\em Department of Physics and Astronomy, University of Southampton, Highfield,
Southampton SO17 1BJ, UK}
\end{center}
\vskip0.7cm  \centerline{\bf ABSTRACT}  
We present viable F-term realizations of the hybrid inflationary scenario
in the context of supergravity addressing at the same time the well-known
problems of the initial conditions and of the adequate suppression of the
inflaton mass. An essential role in our construction is played by ``decoupled"
superheavy fields without superpotential acquiring large vevs due to D-terms
associated with ``anomalous" U(1) gauge symmetries. The naturalness of the
initial conditions is achieved through a ``chaotic" inflation starting at energies
close to the Planck scale and driven by the ``anomalous" D-terms. We discuss
two distinct mechanisms leading to such an early ``chaotic" D-term inflation
which depend on the choice of the K\"ahler potential involving the superheavy
fields. The one relies on a choice of the K\"ahler potential of the $SU(1,1)/U(1)$
K\"ahler manifold of the type encountered in no-scale supergravity whereas the
other employs a more ``conventional" choice for the K\"ahler potential of the
$SU(1,1)/U(1)$ or $SU(2)/U(1)$ K\"ahler manifold but invokes rather specific values
of the Fayet-Iliopoulos $\xi$ term. For such specific values of the $\xi$ term we
exploit the existence of special classical non-oscillatory solutions describing
inflationary expansion at both large and small inflaton field values. In the scenarios
considered the superpotential is linear in the inflaton field associated with the
``observable" inflation on the account of a R-symmetry and the suppression
of the inflaton mass is a result of a cancellation between the always positive
contribution of the ``decoupled" superheavy fields and a negative contribution
originating from the part of the K\"ahler potential involving the inflaton.

\newpage

\setcounter{equation}{0}

\section{Introduction}

Inflation offers an elegant solution to many cosmological problems \cite{linde90}. However, ``natural''
realizations of the inflationary scenario are hard to find. ``New'' and ``chaotic'' inflation \cite{linde90}
invoke a very weakly coupled scalar field, the inflaton, in order to reproduce the observed temperature
fluctuations $\frac{\Delta T}{T}$ \cite{cobe} in the cosmic background radiation. To overcome this
naturalness problem Linde proposed the ``hybrid'' inflationary scenario \cite{hyb} involving a coupled
system of (at least two) scalar fields which manages to produce the temperature fluctuations with natural values
of the coupling constants. This is achieved by exploiting the smallness in Planck scale units
($M_{P}/\sqrt{8\pi }\simeq 2.4355 \times10^{18}$ GeV$=1$ which are adopted throughout our disscusion)
of the false vacuum energy density associated with the phase transition leading to the spontaneous breaking
of a symmetry in the post-Planck era.

The hybrid inflationary scenario, although very simple and attractive, had to face two important challenges,
namely its confrontation with the problem of the initial conditions and its implementation in the context of
global supersymmetry and supergravity. Much effort has been devoted to both these directions of research
which proved that the hybrid inflationary scenario can be naturally realized in supersymmetric theories and
simple extensions of it can also cope with the problem of the initial conditions. The more difficult question,
however, of finding elegant realizations in the context of supergravity with natural initial conditions has not,
in our opinion, been addressed equally successfully. The present work is an attempt in this direction.

The organization of the paper is as follows. In section 2 we review the non-supersymmetric hybrid inflationary
scenario, we state the problem of the initial conditions and discuss possible solutions. Particular emphasis is given
to the necessity of a short  Planck-scale inflationary stage. In section 3 we consider supersymmetric extensions
of the hybrid inflationary scenario. After a brief discussion of hybrid inflation in global supersymmetry we review
the supergravity scenarios proposed in the case that the K\"ahler potential involving the inflaton field is the minimal
or quasi-minimal one making an effort to provide, whenever possible, analytic expressions connecting the scale of
symmetry breaking with the parameters of the model and the data from observations. Later, we also describe an
alternative construction involving K\"ahler potentials very different from the minimal one and ``decoupled" fields
acquiring large vevs through D-terms associated with ``anomalous" $U(1)$ gauge symmetries. In section 4 we
present what we call a ``chaotic" D-term inflation which takes place at energies close to the Planck scale and which
is used to solve the initial condition problem of supergravity hybrid inflation. The mechanism is built in the alternative
construction involving the ``decoupled" fields with large vevs through ``anomalous" D-terms and comes in two
variants depending on the choice of the K\"ahler potential of the ``decoupled" fields. In section 5 we present the
initial conditions that lead to a successful ``observable" inflation according to various scenarios discussed in section
3 after an early ``chaotic" D-term inflation. Finally, in section 6 we briefly present our conclusions.

\setcounter{equation}{0}

\section{Hybrid inflation and the problem of the initial\\ conditions}

The original hybrid inflationary scenario is realized in the simple
model described by the potential \cite{hyb}, \cite{lyth}
\begin{equation}
 \label{hprot}
V(\varphi ,\sigma )=(-\mu ^{2}+\frac{1}{4}\lambda \varphi ^{2})^{2}+
\frac{1}{4}\lambda _{1}\varphi ^{2}\sigma ^{2}+\frac{1}{2}\beta \mu ^{4}\sigma ^{2},
\end{equation}
where $\varphi ,\sigma $ are real scalar fields, $\mu $ is a mass parameter
and $\lambda ,\lambda _{1},\beta $ are real positive constants.
Anticipating supersymmetry we set $\lambda _{1}=\lambda ^{2}$.
Notice that at $\sigma ^{2}=\sigma _{c}^{2}=
2\frac{\mu ^2 }{\lambda}$ the $\sigma$-dependent mass-squared
of $\varphi ,$ $m_{\varphi }^{2}(\sigma )=\lambda(-\mu ^{2} +\frac{1}{2}
\lambda \sigma ^{2}),$ vanishes. Then, $m_{\varphi }^{2}(\sigma )>0$ for
$\sigma ^{2}>\sigma _{c}^{2}$ and the potential at fixed $\sigma $ as a
function of $\varphi $, namely $V_{\sigma }(\varphi )$, has a minimum at $
\varphi =0$ with $V_{\sigma }(0)=\mu ^{4}(1+\frac{1}{2}\beta \sigma ^{2}).$
For $\sigma ^{2}<\sigma _{c}^{2}$ instead, $m_{\varphi }^{2}(\sigma )<0$ and 
$V_{\sigma }(\varphi )$ has a minimum at $\left| \frac{\varphi }{2}\right|
=\left( -\frac{m_{\varphi }^{2}(\sigma )}{\lambda ^{2}}\right) ^{\frac{1}{2}
}.$ Moreover, $\left| \frac{\varphi }{2}\right| =M \equiv \frac{\mu }{\sqrt{\lambda }}
,\sigma =0$ minimizes $V(\varphi ,\sigma ).$

Let us assume that $\frac{2}{\beta }$ $\gg \sigma ^{2}>$ $\sigma _{c}^{2},$
$\varphi =0$ and $\beta \ll 1$.
Then, the potential is dominated by the almost constant
false vacuum energy density, i.e.
$V(0,\sigma )=\mu ^{4}(1+\frac{1}{2}\beta \sigma ^{2}) \simeq \mu ^{4},$
the slow-roll parameters $\epsilon ,\left|\eta\right| \ll 1,$
where $\epsilon \equiv \frac{1}{2}\left(\frac{V^{^{\prime }}}{V}\right) ^{2}
=\frac{1}{2}\beta ^{2}\sigma ^{2}\ll
\beta ,$ $\eta \equiv \frac{V^{^{\prime \prime }}}{V} =\beta ,$
and the universe experiences an inflationary stage with Hubble parameter
$H\simeq \frac{\mu ^{2}}{\sqrt{3}}$. During inflation the motion of the inflaton
field $\sigma $ is governed, in the slow-roll approximation, by the
equation $\frac{d\sigma }{dt}\simeq-\frac{1}{\sqrt{3}}\beta \mu ^{2}\sigma$.
Inflation ends at $\sigma^2 \simeq \sigma_c^2=2M^2$ with a rapid phase transition
towards the true minimum $\left| \frac{\varphi }{2}\right| =\frac{\mu }{\sqrt{\lambda }}, \sigma =0$.
The number of e-foldings for the cosmic time interval $[t_{in},t_{f}]$ during which $\sigma$ varies
between the values $\sigma _{in}$ and $\sigma _{f}$ (with $\sigma _{in}^{2}>\sigma_{f}^{2}$) is
\begin{equation}
\int_{t_{in}}^{t_{f}}Hdt \simeq \beta ^{-1}\ln \frac{\sigma _{in}}{\sigma _{f}}
=N(\sigma _{in})-N(\sigma _{f})
\end{equation}
with $N(\sigma )\equiv \beta ^{-1}\ln \frac{\sigma }{\sigma _{c}}.$  Also the
spectral index of density perturbations $n\simeq1+2\beta $
is slightly larger than $1$ for $\beta \ll 1 .$ 
Assuming that the measured
(quadrupole) anisotropy $\frac{\Delta T}{T}\simeq 6.6\times 10^{-6}$
is dominated by its scalar component
$\left( \frac{\Delta T}{T}\right) _{S}\simeq \left( 12\sqrt{5}\pi 
V^{^{\prime }}\right) ^{-1}V^{\frac{3}{2}}$ (evaluated at $\sigma =\sigma
_{H}=\sigma _{c}e^{\beta N_{H}}$, where $N_{H}\equiv N(\sigma _{H})\simeq
50-60$ is the number of e-foldings of the ``observable''
inflation), we have 
\begin{equation}
 \label{hprot1}
\lambda M=12 \sqrt{10}\pi \frac{\Delta T}{T}{\beta }e^{\beta N_{H}}
\simeq 7.87\times 10^{-4}{\beta }e^{\beta N_{H}}.
\end{equation}
Taking for $M$ the MSSM scale
\begin{equation}
M=\frac{M_{X}}{g_{G}}\simeq1.17\times 10^{-2},
\end{equation}
where $M_{X} \simeq 2 \times 10^{16}$ GeV
is the mass acquired by the superheavy gauge bosons and $g_{G}\simeq 0.7$ the
unified gauge coupling constant, we obtain
\begin{equation}
\lambda \simeq 0.067 \beta e^{\beta N_{H}}.
\end{equation}
For $N_{H}=55$ and $\beta\simeq (1 - 3)\times 10^{-2}$ or
equivalently $n\simeq 1.02 - 1.06$
(i.e. a moderately blue spectrum of density perturbations)
we have $\lambda\simeq (1 - 10) \times 10^{-3},$ 
$\mu\simeq (4 - 12) \times 10^{-4}$ and $\sigma_{H} \simeq (3 - 9) \times 10^{-2}$.
An almost scale invariant spectrum can be obtained for $\beta\ll10^{-2}$
but then, insisting on the MSSM value for $M$, 
$\lambda\simeq0.067\beta$ becomes unnaturally small.

At this point we should remark that $\mu$ cannot be arbitrarily large since there is an upper bound on the energy
density scale $V_{infl}^{\frac{1}{4}}\simeq V_{\sigma _{H}}^{\frac{1}{4}}\simeq \mu$ where the ``observable''
inflation begins. By exploiting the fact that the tensor component
$\left( \frac{\Delta T}{T}\right) _{T}^2\simeq \left( 720\pi ^{2}\right)^{-1}6.9V_{\sigma _{H}}$
of $\left(\frac{\Delta T}{T}\right)^2$ satisfies $\left( \frac{\Delta T}{T}\right) _{T}^{2}\le
\left( \frac{\Delta T}{T}\right) ^{2}$ we immediately derive the bound
\begin{equation}
 \label{tensor}
V_{infl}^{\frac{1}{4}}\simeq V_{\sigma _{H}}^{\frac{1}{4}}\simeq \mu 
\lesssim 1.46\times 10^{-2}.
\end{equation}
Our assumption that the scalar component dominates the quadrupole anisotropy
is equivalent to $\mu$ having a value well below the above bound.

The above discussion of the hybrid inflationary scenario is certainly simplified
since it is restricted to field values along the inflationary trajectory $\varphi=0$.
The naturalness of the scenario, however, depends on the existence of 
field values which although initially are far from the inflationary trajectory
they approach it during the subsequent evolution \cite{in}, \cite{tet}.
We assume that the energy density 
$\rho$ of the universe is dominated by $V(\varphi ,\sigma )$.
Let us start away from the inflationary trajectory and choose the energy
density $\rho _{0}$ to satisfy the relation $\mu ^{4}\ll \rho _{0}\lesssim 1.$
Moreover, we assume that $\varphi ^{2}$ starts somewhat below $\sigma ^{2}$
(i.e. $\varphi_{0} ^{2} \lesssim \sigma_{0} ^{2}$).
Then, the relevant term in $V$ for our discussion is the term
$\frac{1}{4}\lambda ^{2}\varphi ^{2}\sigma^{2}.$
We would like $\varphi $ to oscillate from the beginning as a
massive field due to its coupling to $\sigma $ and quickly become very close
to zero. In contrast, $\sigma ^{2}$ should stay considerably larger than
$\sigma _{c}^{2}$. Thus, for $\mu ^{4}\ll \rho \leq \rho _{0}\lesssim 1$ it is
required that $\frac{4}{9}\frac{m_{\sigma }^{2}}{H^{2}}\ll 1\ll \frac{4}{9}
\frac{m_{\varphi }^{2}}{H^{2}}$ or $\varphi ^{2}\ll \frac{8}{3}\ll \sigma
^{2}.$ When $\rho \sim \mu ^{4}$, instead, $\left|\sigma\right|$ remains larger
than $\left|\sigma_{c}\right|$ provided 
$\frac{4}{3}\frac{m_{\sigma }^{2}}{\mu 4}\lesssim 1$
or $\varphi ^{2}\lesssim \frac{3}{2}\frac{\mu ^{4}}{\lambda ^{2}}$.
If we allow $\left| \sigma _{0}\right| \gg 1$, $\left|
\varphi _{0}\right| $ does not have to be very small. For example, with
$\beta \simeq (1 - 3)\times 10^{-2}$ we could have
$\left| \sigma _{0}\right| \simeq 4.5$, $\left|
\varphi _{0}\right| \simeq 1$. If, instead, we insist that $\left| \sigma
_{0}\right| <1$ we are forced to start very close to the inflationary
trajectory $\rho_{0} \simeq \mu^{4} \ll 1$ and severely fine tune the starting
field configuration ($\left| \sigma_{c}\right| \ll \left| \sigma_{0}\right|<1, $
$\left|\varphi _{0}\right| \lesssim \frac{\mu ^{2}}{\lambda }\ll 1$) \cite{tet}.

This severe fine tuning becomes more disturbing since the
field configuration at the assumed onset of inflation,
where $H=H_{infl}$, should be homogeneous over dinstances
$\sim H_{infl}^{-1}$. Notice that $H_{infl}^{-1}$ is larger than the Hubble
distance at the end of the Planck era ($\rho =$ $\rho _{in}\simeq 1$) as
expanded (according to the expansion law $R\sim \rho ^{-\frac{1}{3\gamma }},$
where $R$ is the scale factor of the universe) till the assumed onset of
inflation (at $\rho =\rho _{_{infl}}$) by a factor $\frac{H_{infl}^{-1}}{
H_{in}^{-1}}\left( \frac{\rho _{infl}}{\rho _{in}}\right) ^{\frac{1}{3\gamma }
}=\left( \frac{\rho _{in}}{\rho _{infl}}\right) ^{\frac{3\gamma -2}{6\gamma }
}\gg 1,\ $if$\ \ \gamma \gtrsim 1.$ Therefore, in order for (any type of)
inflation to start at an energy density scale $\rho _{infl}^{\frac{1
}{4}}\simeq $ $V_{infl}^{\frac{1}{4}}\ll 1$, the initial field configuration
at $\rho =\rho _{in}\simeq 1$ (where initial conditions should be set)
must be very homogeneous over distances $\sim \left( V_{infl}^{-\frac{1}{4}
}\right) ^{2\frac{3\gamma -2}{3\gamma }}\gg 1$. Such a homogeneity is hard to
understand unless a short period of inflation took place at
$\rho \sim 1$ \cite{double} with a number of e-foldings
$\gtrsim 2\frac{3\gamma -2}{3\gamma }\ln \left(V_{infl}^{-\frac{1}{4}}\right)$.
An early inflationary stage might also eliminate the requirement of severe fine
tuning of the field configuration at $\rho=\rho_{in}$ since,
in addition to the homogenization of space, it could alter the dynamics
during the early stages of the evolution of the universe.

An inflation taking place at an energy density $\rho_{1}\gg\rho_{infl}$, however,
although eliminates existing inhomogeneities it generates new ones due to
quantum fluctuations. These fluctuations are $\sim \frac{H_{1}}{2\pi }$ for
massless fields and generate inhomogeneities over distances $\sim H_{1}^{-1}$
resulting in a gradient energy density $\sim \frac{H_{1}^{4}}{4\pi ^{2}}=\frac{
\rho_{1} ^{2}}{36\pi ^{2}}$ which falls with the expansion only like
$R^{-2}\sim \rho ^{\frac{2}{3\gamma }}.$
The size of this gradient energy density when $\rho $ falls to $\rho
_{infl}\simeq V_{infl}$ should be smaller than $V_{infl}.$
This gives an upper bound on the energy density $\rho_{1}$
(towards the end) of the first stage of inflation
\begin{equation}
  \label{bound1}
\rho _{1}\lesssim \left( 6\pi \right) ^{\frac{3\gamma }{3\gamma -1}}\left(
V_{infl}^{\frac{1}{4}}\right) ^{2\frac{3\gamma -2}{3\gamma -1}} \qquad
\left(\gamma \gtrsim 1 \right)
\end{equation}
which is somewhat lower than unity and decreases with $V_{infl}^{\frac{1}{4}}$.

Such an early inflationary stage can be easily incorporated into the hybrid
model \cite{double}. In particular, if we allow field values considerably
larger than unity (e.g. $\left| \varphi _{0}\right| =\left| \sigma _{0}\right|
\gtrsim 10$) the original model gives rise
to an early chaotic-type inflationary stage at $\rho= \rho_{0} \sim \rho_{in}$
which takes care of the initial condition problem.

It is important to realize that the above arguments in favor of
an early inflationary stage as a solution to the initial condition
problem of the ``observable" inflation are too general to be 
applicable to the hybrid inflationary scenario only. As an example
of the usefulness of an early inflation in other cases as well we mention
the ``new" inflationary scenario whose serious initial condition problem
may be solved by a stage of ``pre-inflation" \cite{pre}.

\setcounter{equation}{0}

\section{Hybrid inflation in global supersymmetry and\\ supergravity}

Let us first treat supersymmetry as only global. We consider a model with gauge group $G$
which breaks spontaneously at a scale $M.$ The symmetry breaking of $G$ is achieved through
a superpotential which includes the terms
\begin{equation}
  \label{sup}
W=S(-\mu ^{2}+\lambda \Phi \bar{\Phi}).
\end{equation}
Here $\Phi ,\bar{\Phi}$ is a conjugate pair of left-handed superfields which belong to a non-trivial
$N_d-$dimensional representation of $G$ and break it by their vevs, $S$ is a gauge singlet left-handed
superfield, $\mu $ is a mass scale related to $M$ and $\lambda $ a real and positive coupling constant.
The superpotential terms in $W$ are the dominant couplings involving the superfields $S$, $\Phi $,
$\bar{\Phi}$ which are consistent with a continuous R-symmetry under which $W\to e^{i\vartheta }W$,
$S\to e^{i\vartheta }S$, $\Phi\bar{\Phi} \to \Phi\bar{\Phi}$ \cite{dss}. The potential obtained from $W$ is
\begin{equation}
V=\left| -\mu ^{2}+\lambda \Phi \bar{\Phi}\right| ^{2}+\lambda^{2}\left|S\right|
^{2}(\left| \Phi \right| ^{2}+\left| \bar{\Phi}\right| ^{2})+V_D,
\end{equation}
where the scalar components of the superfields are denoted by the same symbols as the corresponding
superfields and $V_D$ is the D-term contribution. The supersymmetric minimum $S=0,$ $\Phi \bar{\Phi}=$
$\frac{\mu ^{2}}{\lambda}=M^2,$ $\left| \Phi \right|=\left| \bar{\Phi}\right| $ lies on the D-flat direction
$\Phi =\bar{\Phi}^{*}$. By appropriate gauge and R-transformations on this D-flat direction we can bring
the complex $\Phi $, $\bar{\Phi}$, $S$ fields on the real axis, i.e. $\Phi =\bar{\Phi}\equiv \frac{1}{2}\varphi$,
$S\equiv \frac{1}{\sqrt{2}}\sigma$, where $\varphi$ and $\sigma$ are real scalar fields. Then, the potential
acquires the form of Eq.\  (\ref{hprot}) with $\lambda _{1}=\lambda ^{2}$ and $\beta=0$.

A tiny mass-squared $m_{\sigma}^2 = \beta \mu^4 \sim 1$ Te$V^2$ can be generated for $\sigma$
as a result of soft supersymmetry breaking. The parameters of the resulting hybrid inflationary scenario 
with the implicit assumption that the slope $V^{\prime}$ is dominated by the mass-squared term \cite{lyth}
satisfy the relation $\lambda M^{\frac{5}{3}}\sim (5-6) \times10^{-12}.$ Consequently, the symmetry breaking
scale $M$ must be smaller than the MSSM scale $\sim 10^{-2}$ unless $\lambda$ is unnaturally small.
(Actually, as it turns out from the calculations that we present later, there is also the bound
$\lambda M^4 \gtrsim 5\times 10^{-20}$ in such a scenario which combined with the relation determining
the value of $\lambda M^{\frac{5}{3}}$ gives $M \gtrsim 3.5 \times 10^{-4}$ and $\lambda \lesssim 3.5 \times 10^{-6}$.
Therefore, domination of the tiny mass-squared term always results in an unnaturally small coupling.)

Alternatively, one can adopt the point of view that the dominant contribution to the slope $V^{\prime}$ is
provided by radiative corrections \cite{dss} through the simplified expression
\begin{equation}
  \label{sl1}
V_{rad}^{\prime} \simeq
N_d\left(\frac{\lambda}{2\pi}\right)^2\frac{\mu^4}{2\sigma}.
\end{equation}
Then, on the condition that
\begin{equation} 
\left|\eta_c^r\right|\equiv\left|\frac{V_{rad}^{\prime \prime}}{V}\right|_{\sigma^2=\sigma_c^2=2M^2}
=\left(\frac{\sqrt{N_d}\lambda }{4\pi M}\right)^2
\end{equation}
satisfies $\left|\eta_c^r\right|\ge 1$ (i.e. inflation ends due to large radiative corrections), we obtain
\begin{equation}
 \label{s1}
M^2=\sqrt{\frac{45N_d}{N_H+\frac{1}{2}}}\frac{\Delta T}{T}.
\end{equation}
If, instead, $\left|\eta_c^r\right| \le 1$ (i.e. inflation ends through the waterfall mechanism), we obtain
\begin{equation}
  \label{s4}
\lambda = \frac{4\pi M}{\sqrt{N_d}}\frac{1}{\sqrt{2}}\left(\frac{45N_d(\frac{\Delta T}{T})^2}{M^4} - N_H\right)^{-\frac{1}{2}}.
\end{equation}
Here the scale $M$ is bounded from above by the value given by Eq.\  (\ref{s1}). $M$ is always smaller
than the MSSM scale unless $N_d$ is extremely large. The value of the inflaton field $\sigma$ when our
present horizon crossed outside the inflationary horizon is always given by
\begin{equation}
  \label{s2}
\sigma_{H}^2=N_d\left(\frac{\lambda}{2\pi}\right)^2
\left(N_H+\frac{1}{2}max\left\{1, \left|\eta_c^r\right|^{-1}\right\}
\right)
\end{equation}
and the spectral index at the scale corresponding to our present horizon is
\begin{equation}
  \label{s3}
n_H\simeq 1-\frac{1}{N_H+\frac{1}{2}max\left\{1, \left|\eta_c^r\right|^{-1}\right\}}.
\end{equation}

An attempt to obtain the MSSM value for the scale $M\simeq 1.17\times 10^{-2}$ by replacing the term
$\lambda\Phi \bar{\Phi}$ in $W$ by $\lambda (\Phi \bar{\Phi})^2$ through a $Z_2$ symmetry gave rise to the
so-called ``smooth" hybrid inflation \cite{smooth}. An additional advantage of this scenario is that it avoids
the formation of unwanted topological defects since the phase transition leading to the termination of the
inflationary stage takes place ``smoothly".

Supersymmetry cannot, of course, remain just global. Incorporation of supergravity is, however, by no
means a trivial task. A very well-known difficulty in this connection is the one of keeping the slow-roll parameter
$\left|\eta\right|$ small. Indeed, in the very common case that the potential during inflation is dominated by the
F-term, supergravity tends to give a large mass to almost all fields, thereby eliminating most candidate inflatons
\cite{lyth}, \cite{st}. This can be easily seen by considering the F-term potential in supergravity
\begin{equation}
V_{F}=e^{K}(\cdots ),
\end{equation}
where $K$ is the K\"ahler potential. Let us assume that our candidate inflaton field $S$ is canonically normalized
for $\left| S\right| ^{2}\ll 1$ and $K$ admits an expansion $K=\left| S\right|^{2}+\ldots $ . Then,
\begin{equation}
m_{S}^{2}=K_{SS^*}V_{F}+\ldots
=\left( 1+\ldots \right) V_{F}+\ldots=V_{F}+\ldots \ ,
\end{equation}
where the subscript $S$ ($S^*$) denotes partial differentiation with respect to $S$ ($S^*$).
Thus, during inflation, no matter how small $V_{infl}\simeq V_{F_{infl}}$ is, there is always
a contribution to $m_{S}^{2}$  $\simeq V_{infl}$ or a contribution $\simeq 1$ to the slow-roll parameter
$\left|\eta\right|.$ There could very well exist other contributions to $\eta$ partially cancelling the one just
described but their existence will depend on the details of the model. Therefore, it seems that in the context
of supergravity it is fairly easy to add to the potential of the hybrid model a sizeable mass-squared term for
the inflaton $\sigma$ but rather difficult to understand why $\beta\equiv \frac{m_{\sigma}^{2}}{\mu ^{4}} \ll 1$.

In order to investigate the effect that supergravity has on the simple globally supersymmetric hybrid model
discussed above we restrict ourselves to the inflationary trajectory ($\Phi $ $=\bar{\Phi}=0$) in which case the
superpotential simplifies to
\begin{equation}
W=-\mu ^{2}S
\end{equation}
involving just the gauge singlet superfield $S$ which enters only linearly.
Also the K\"ahler potential becomes a function of just $\left|S\right|^2$
on the account of the same R-symmetry which guarantees the linearity of $W$ in $S$.
Then, a straightforward calculation shows that the mass-squared term generated for
the inflaton corresponds to a parameter $\beta$ given by
\begin{equation}
\beta=-(K_{SSS^{*}S^{*}})_{S=0}.
\end{equation}
Notice that the unwanted contribution $(K_{SS^{*}})_{S=0}$ to $\beta$
``miraculously" cancels out. As a consequence $\beta$ could now
have any value since the quantity $K_{SSS^{*}S^{*}}$ does not seem
to be constrained by any general argument. In particular, the choice of the
minimal K\"ahler potential $K=\left| S\right| ^{2}\ $leading to canonical kinetic terms
for $\sigma $ gives rise to a ``canonical'' potential $V_{can}$ \cite{lyth}, \cite{pan},
\cite{linde97}
\begin{equation}
  \label{can}
\frac{V_{can}}{\mu ^{4}}
=\left(1-\frac{\sigma^2}{2}+\frac{\sigma^4}{4}\right)e^{\frac{\sigma^2}{2}}
=1+\sum_{k=1}^{\infty }\frac{(k-1)^{2}}{2^{k}k!}\sigma^{2k}
\end{equation}
without quadratic term ($\beta=0$). Small deviations from the
minimal form of the K\"ahler potential generate a ``small" mass-squared
for $\sigma$ ($\beta\ll1$) without, of course, altering appreciably the
coefficients of the higher powers of $\sigma^2$ in the series \cite{pan1}.

The potential $V_{can}$ leads to acceptable hybrid inflation with
a scale $M$ given by the MSSM value either in the context of the
simple model of Eq.\  (\ref{sup}) with $N_d=1$ and a very small coupling
$\lambda\simeq 10^{-5}$ or in the context of models involving
additional symmetry breaking scales \cite{pan}.
In both cases supergravity dominates over radiative corrections.
An acceptable inflationary scenario in the context of the
model of Eq.\  (\ref{sup}) and a minimal K\"ahler potential is also feasible
if radiative corrections are allowed to play a dominant role
at least towards the end of inflation \cite{linde97}. In this last case, however,
one has to either give up the MSSM value for the scale $M$ or
allow $N_d$ values considerably larger than unity. 
 
To demonstrate the above statements we perform a simplified calculation, with $V\simeq \mu^4$ and 
\begin{equation}
  \label{rad1}
V^{\prime} \simeq
\left(N_d\left(\frac{\lambda}{2\pi}\right)^2+c\sigma^4\right)\frac{\mu^4}{2\sigma}
\end{equation}
assuming that inflation takes place at $\sigma^2  \ll 1$ and the dominant supergravity effect is due to
the quartic self-coupling of the inflaton $\frac{c}{8}\mu^4\sigma^4$. Then, in the context of the model
of Eq.\  (\ref{sup}) we obtain
\begin{equation}
  \label{rad2}
M^2=\sqrt{\frac{45N_d}{N_H}}\frac{\Delta T}{T}
\sqrt{\frac{\phi}{\tan\phi}}\left(1+\tan^2\phi\right)
f_1^{\frac{3}{2}}f_2
\end{equation}
with
\begin{equation}
  \label{rad3}
\sigma_{H}^{2}=\frac{\phi \tan\phi}{cN_H}f_1
\end{equation}
and
\begin{equation}
  \label{rad4}
n_H\simeq 1+\frac{1}{N_H}\frac{\phi}{\tan\phi}\left(3 \tan^2\phi - f_1^{-2}\right)f_1,
\end{equation}
where $\phi=N_H\sqrt{cN_d}\frac{\lambda}{2\pi}<\frac{\pi}{2}$,
$f_1=\left(1+\frac{\omega}{\tan\phi}\right)(1 - \omega \tan\phi)^{-1}$,
$f_2=(1 + \omega^2)\left(1+\frac{\omega}{ \tan\phi}\right)^{-2}$
and
$\omega = \frac{\phi}{2N_H} max\left\{1, \frac{4cN_H^2M^2}{\phi^2}\right\}$.
For minimal supergravity $c=1$. Taking the limit $c\to 0$ we recover, 
as expected, the relations of the radiative supersymmetric scenario.

A careful investigation reveals that only for values of the coupling $\lambda$
in a certain interval there exist solutions for $M$. Actually, there exist two 
solutions which coincide at the lower endpoint of this interval where
$\lambda$ takes the value
\begin{equation}
\lambda_{min} \simeq 640\sqrt{3}\pi c^{\frac{3}{2}}
N_d^{\frac{1}{2}}\left(\frac{\Delta T}{T}\right)^2
\end{equation}
with the corresponding value of the double solution for $M$ being
\begin{equation}
M_{d}^2 \simeq 480cN_d \left(\frac{\Delta T}{T}\right)^2.
\end{equation} 
For the upper endpoint $\lambda_{max}$ we have 
$\lambda_{max} \lesssim min \left\{ \frac{\pi^2}{\sqrt{cN_d}N_H} , \frac{2}{\sqrt{N_dN_H}} \right\}$.
Moreover, only solutions with $\sqrt{\lambda}M=\mu$ well below the bound of Eq.\  (\ref{tensor}) are acceptable.
Out of the two solutions in the allowed interval the one with the larger value for the scale $M$ owes its existence
to supergravity and tends to infinity as $c\to 0$ at fixed $\lambda$. The second solution survives the $c\to 0$
limit in which case it reduces to a solution describing the radiative supersymmetric scenario. In the following
we refer to them as the ``supergravity" and the ``radiative" solution, respectively.

The ``supergravity" solution \cite{pan} can be approximated by
\begin{equation}
\lambda M^2 =\mu^2\simeq 6\sqrt{5}\pi \frac{\Delta T}{T}c
\left(\frac{\sigma_c^2}{1-cN_H\sigma_c^2}\right)^{\frac{3}{2}}
\end{equation}
with
\begin{equation}
\sigma_H^2\simeq \frac{\sigma_c^2}{1-cN_H\sigma_c^2} 
\end{equation}
and
\begin{equation}
n_H \simeq 1+ \frac{3c\sigma_c^2}{1-cN_H\sigma_c^2}
\end{equation}
assuming $320\sqrt{3}c^2N_dN_H\left(\frac{\Delta T}{T}\right)^2 \ll \phi \ll cN_H\sigma_c^2 < 1$.
We observe that the spectrum of density perturbations is blue ($n_H > 1$). In the model of
Eq.\  (\ref{sup}), where $\sigma_c^2=2M^2$, scales $M\sim 10^{-2}$ are obtainable but at
$\lambda \ll1$ (or $\mu^2 \ll M^2$). Larger scales close to $\frac{1}{\sqrt{2cN_H}}$ are, instead,
favored. By extending the model of Eq.\  (\ref{sup}) to accommodate additional symmetry breaking
scales we can raise $\sigma_c^2$ above $2M^2$ and with increasing $\mu^2$ and
${\sigma_H^2}/{\sigma_c^2}$ obtain blue perturbation spectra with $n_H$ considerably larger
than 1 even if $M\sim 10^{-2}$ \cite{pan}.

The more popular ``radiative" solution \cite{linde97} can be approximated by setting
\begin{equation}
f_{1,2}\simeq 1.
\end{equation}
This is justified provided $M_{d} \ll M$ and $8\sqrt{cN_H}M\lesssim \phi \lesssim \frac{3\pi}{8}$.
The lower bound in the above inequality for $\phi$ translates into $\left|\eta_c^r\right|\gtrsim \frac{16}{N_H}$.
Thus, radiative corrections should be neither extremely strong nor extremely weak. We see that
with radiative corrections becoming stronger $\left|\sigma_H\right|$ increases. As a result the
spectrum changes gradually from red to blue and larger values of the scale $M$ become possible.
However, the MSSM value for $M$ is obtainable only if $N_d$ is considerably larger than unity.
For example, for $\phi \simeq 1.12293$ (1.1471), $N_d=32$ (27), $N_H=55$ and $c=1$, and taking
into account the small effect of the factors $f_1$, $f_2$ as well, we obtain (the MSSM scale)
$M\simeq 1.173 \times 10^{-2}$ with $\lambda \simeq 2.27 \times 10^{-2}$ ($2.52 \times 10^{-2}$),
$\sigma_H \simeq 0.21$ (0.22) and $n_H\simeq 1.12$ (1.13). Notice that for such a choice inflation
ends through the waterfall mechanism.      

The addition of a quadratic term $\frac{1}{2}\beta\mu^4\sigma^2$ in 
$V_{can}$ changes the picture radically \cite{pan1}. Now, as we shall see
shortly, the MSSM value for the scale $M$ becomes readily obtainable in the
simple model of Eq.\  (\ref{sup}) with $N_d=1$ and reasonable values
of the parameters.

In the presence of an appreciable mass term the quartic coupling ceases being the dominant effect
of supergravity for $\sigma^2 \ll1$. Then, a simplified discussion which ignores the quartic
and higher order self-couplings of the inflaton and assumes that $V\simeq \mu^4$ and 
\begin{equation}
V^{\prime} \simeq
\left(N_d\left(\frac{\lambda}{2\pi}\right)^2+2\beta\sigma^2\right)\frac{\mu^4}{2\sigma}
\end{equation}
could be very illuminating. In the case that inflation ends because radiative corrections
become strong, i.e. $\left|\eta_{c}^r\right| \ge 1$, we find that the scale $M$ is given by
\begin{equation}
   \label{mass1}
M^2=M_{rad}^2\equiv
\sqrt{\frac{45N_d}{\left(\frac{1-e^{-2\beta N_H}}{2\beta}+\frac{1}{2}\right)}}
\frac{\Delta T}{T} (1+\beta)e^{\beta N_H}.
\end{equation}
If, instead, inflation ends through the waterfall mechanism, i.e. $\left|\eta_{c}^r\right| \le 1$, we obtain
\begin{equation}
   \label{mass2}
\left(\frac{12\sqrt{10}\pi{\frac{\Delta T}{T}}{ \beta}{e^{\beta N_H}}}
{\lambda M}\right)^2=
\frac{1}{2}\left(1+
s\sqrt{1 - {\frac{360N_d(\frac{\Delta T}{T})^2}{M^4}}\beta}\right)
-{\frac{90N_d(\frac{\Delta T}{T})^2}{M^4}}\beta e^{2\beta N_H},
\end{equation}
where $s=\pm1$. The solution with $s=+1$ is acceptable provided
$M^2 \ge \sqrt{360N_d\beta}\frac{\Delta T}{T}$ for $(1 +\beta)e^{2\beta N_H} \le 2$ or
$M^2 \ge M_{rad}^2$ for $(1 +\beta)e^{2\beta N_H} > 2$. The solution with $s= - 1$ is acceptable only if
$M_{rad}^2 \ge M^2 \ge  \sqrt{360N_d\beta}\frac{\Delta T}{T}$ and $(1 +\beta)e^{2\beta N_H} \le 2$. Always
\begin{equation}
  \label{mass3}
\sigma_H^2=2M^2e^{2\beta N_H}\left(max\left\{1, \left|\eta_c^r\right|\right\}
+\left|\eta_c^r\right| \left(1-e^{-2\beta N_H}\right) \beta^{-1}\right).
 \end{equation}  

Radiative corrections dominate over the mass term in $V^{\prime}$ for $\sigma^2 < \sigma_H^2$
(i.e. throughout inflation) when $(1 +\beta)e^{2\beta N_H} \le 2$ (i.e. $\beta \lesssim \frac{\ln2}{2N_H+1}$)
and inflation ends either through strong radiative corrections or through the waterfall mechanism
provided that, in the latter case, the solution with $s= - 1$ is chosen. In the above cases the $\beta \to 0$
limit exists and Eq.\  (\ref{s1}) or Eq.\  (\ref{s4}) is recovered depending on whether $\left|\eta_{c}^r\right|$
is larger or smaller than unity. In such radiative-correction-dominated scenarios the scale $M$
has both an upper and a lower $\beta$-dependent bound. In particular, the upper bound $M_{rad}$
cannot exceed a value $\simeq 3.15 \times 10^{-3}N_d^{\frac{1}{4}}$ for $N_H=55$.
On the other hand the choice $s=+1$ in Eq.\  (\ref{mass2}) offers a scale $M$ bounded only from below.
Moreover, when $M^4 \gg 360N_d\left(\frac{\Delta T}{T}\right)^2\beta$ the r.h.s. of Eq.\  (\ref{mass2})
with $s=+1$ is close to unity and Eq.\  (\ref{hprot1}) relating the parameters of the prototype hybrid
inflationary scenario is recovered. Thus, the MSSM value for the scale $M$ is now obtainable for
$\beta\simeq (1 - 3) \times 10^{-2}$ and $N_d=1$ with $\lambda$, $\sigma_H$ and $n_H$ having values
not very different from the ones they take when radiative corrections are neglected.

A quartic self-coupling of the inflaton $\frac{c}{8}\mu^4\sigma^4$ and higher order such terms
originating from supergravity will certainly affect the above discussion to some extent.
An estimate of their effects for the case we are interested in, namely $M\simeq 1.173 \times 10^{-2}$
and $\beta \simeq (1 - 3) \times 10^{-2}$, can be inferred from the value of the ratio $c\sigma_H^2/2\beta$
which for $c\simeq1$ varies between approximately $0.04$ and $0.15$. The impact of such supergravity
terms on the value of $n_H - 1$ is, however, 3 times as large. Thus, for $\beta=0.03$ and $c\simeq 1$
we expect $n_H - 1 \simeq 0.09$, instead of $n_H -1 \simeq 2\beta=0.06$ obtained with the quadratic term
alone, indicating that $\beta$ should not assume a much larger value. Clearly, a smaller $c$ could allow
a larger $\beta$.

Our discussion so far seems to indicate that on the assumption that all fields but the inflaton
$S$ play absolutely no role during inflation the only potential source of inflaton mass is the 
next-to-leading term in the expansion of the K\"ahler potential in powers of $\left| S\right|^{2}$
which must have a small and negative coefficient. There could, however, exist superheavy fields
which are $G-$ singlets and do not contribute to the superpotential. Although such fields could
naively be considered as ``decoupled" they actually contribute to the mass-squared of $\sigma$
if they acquire large vevs. These new contributions to the parameter $\beta$, which turn out to be
always positive, could be disastrous if $-(K_{SSS^{*}S^{*}})_{S=0}>0$ \cite{lr} but very useful in
the opposite case since they could give rise to interesting cancellations \cite{costas}.

To illustrate the above statements we consider a $G-$singlet chiral superfield $Z$ which does not enter the
superpotential at all because, for instance, it has non-zero charge, let us say $-1,$ under an ``anomalous''
$U(1)$ gauge symmetry and there are no superfields with a $U(1)$ charge which cannot be safely ignored.
Also let us assume that $K=K_{1}(\left| S\right|^{2})+K_{2}(\left| Z\right| ^{2}).$ We rename the parameters
$\mu $ and $\lambda$ in $W$ as $\mu ^{^{\prime }}$ and $\lambda ^{^{\prime }}$ and we assume that
${\mu^{\prime}}^2 \ll \xi$, where $\xi >0$ is the Fayet-Iliopoulos term entering the D-term 
\begin{equation}
V_D=\frac{g^{2}}{2}\left( {K_Z}Z-\xi \right) ^{2}
\end{equation}
of the ``anomalous'' $U(1)$ gauge symmetry which has coupling $g$. Minimization of the scalar potential
(with $\Phi $ $=\bar{\Phi}=0$) with respect to $Z$, for fixed $\left|S\right| ^{2}$ not very large and away
from possible singularities, essentially amounts to minimizing the D-term since the F-term is proportional to
${\mu^{\prime}}^4 \ll \xi^2$. As a result of such a minimization $\left| Z\right|^{2}$ acquires a practically
$\left|S\right|$-independent vev $ v^{2}$ typically $\sim \xi $. It is reasonable to expect that
$\left| Z\right|^{2} \simeq v^2$ during inflation. Then, after absorbing the factor $e^{K_{2}(v^{2})}$ appearing
in the F-term potential in the reintroduced parameters $\mu =\mu ^{^{\prime}}e^{K_{2}(v^{2})/4}$ and
$\lambda =\lambda ^{^{\prime }}e^{K_{2}(v^{2})/2}$, it is not difficult to recognize that the false vacuum energy
density now is $\mu^4$, the symmetry breaking scale remains $M=\frac{\mu }{\sqrt{\lambda }}=
\frac{\mu ^{^{\prime }}}{\sqrt{\lambda ^{^{\prime }}}}$ and the parameter $\beta$ becomes
\begin{equation}
  \label{beta1}
\beta=-(K_{SSS^{*}S^{*}})_{S=0}+\left(\left|K_Z\right|^2
\left(K_{ZZ^*}\right)^{-1}\right)_{\left| Z\right| =v}.
\end{equation}
As already mentioned the contribution of $Z$ to $\beta$ is positive. Notice that the ``decoupled" $Z-$type
fields affect the potential during inflation essentially only through the parameters $\mu$ and $\beta$. 

The above mechanism could be readily applied in order to supplement the potential $V_{can}$ with
a small mass-squared term without departing from the minimal K\"ahler $K_1(\left|S\right|^2)=\left|S\right|^2$.
This results in the addition to $V_{can}$ of a perturbation 
\begin{equation}
  \label{cand}
\delta V_{can}=\frac{1}{2}\beta \mu^4{\sigma^2}e^{\frac{\sigma^2}{2}}
=\beta \mu^4 \sum_{k=1}^{\infty }\frac{\sigma^{2k}}{2^{k}(k-1)!}.
\end{equation}

The most interesting case, however, concerns K\"ahler potentials $K_1(\left|S\right|^2)$
with $-({K_1}_{SSS^{*}S^{*}})_{S=0}<0$ which offer the possibility of suppressing the inflaton mass.
A class of such K\"ahler potentials is given by 
\begin{equation}
    \label{KS}
K_{1}(\left| S\right| ^{2})=-N \ln\left( 1-\frac{\left| S\right| ^{2}}{N}
\right) \qquad \left( \left| S\right| ^{2}<N\right) ,
\end{equation}
where $N$ is an integer. The corresponding K\"ahler manifold is the coset space $SU(1,1)/U(1)$
with constant scalar curvature $2/N$ and $({K_1}_{SSS^{*}S^{*}})_{S=0}=2/N$. For this choice
the canonically normalized inflaton $\sigma$, defined by 
\begin{equation}
ReS\equiv \sqrt{N} \tanh{\frac{\sigma}{\sqrt{2N}}},
\end{equation}
acquires a potential $V$ along the inflationary trajectory which in terms of the variable
$x=\cosh^2\frac{\sigma}{\sqrt{2N}}$ is given by
\begin{eqnarray}
\frac{V}{\mu^4}=1+(N-1)(N-2)\left(x^N-1-(2N-1)\frac{x^{N-1} - 1}{N-1}
+(N-1)\frac{x^{N-2} - 1}{N-2}\right) \nonumber\\
+\beta N(x-1)x^{N-1},
\end{eqnarray}
with
\begin{equation}
\beta= - \frac{2}{N} + \sum_i \left(\left|K_{Z_i}\right|^2
\left(K_{Z_iZ_i^*}\right)^{-1}\right)_{\left| Z_i\right| =v_i}.
\end{equation}
Here we allow more than one $Z-$type fields $Z_i$. Notice that for $N=1, 2$ the slope $V^{\prime}$
in the absence of radiative corrections is proportional to $\beta$ meaning that suppression of the
inflaton mass entails suppression of all supergravity corrections to the classical potential. In particular,
$\beta=0$ amounts to a completely flat classical inflationary potential. Moreover, the slope of the
inflationary trajectory due to radiative corrections $V_{rad}^{\prime}$, when expressed in terms of the
canonically normalized inflaton field, becomes
\begin{equation}
  \label{sl2}
V_{rad}^{\prime}\simeq N_d\left(\frac{\lambda}{2 \pi}\right)^2
{\frac{\mu^4}{\sqrt{2N}\sinh\left({\sqrt{\frac{2}{N}}}{\sigma}\right)}}.
\end{equation} 

Expanding in a power series we obtain 
\begin{eqnarray}
\frac{V}{\mu^4}=1+\left(1-\frac{1}{N}\right)\left(1-\frac{2}{N}\right)
\left\{{\frac{\sigma^4}{8}}+\left(1-\frac{1}{N}\right){\frac{\sigma^6}{12}}+
\left(1-\frac{23}{9N}+\frac{8}{5N^2}\right){\frac{3\sigma^8}{128}}
+\dots    \right\} \nonumber\\
+\beta\left\{{\frac{\sigma^2}{2}}+\left(1-\frac{2}{3N}\right){\frac{\sigma^4}{4}}+
\left(1-\frac{5}{3N}+\frac{34}{45N^2}\right){\frac{\sigma^6}{16}}
+\dots    \right\}
\end{eqnarray} 
for V and
\begin{equation}
V_{rad}^{\prime}\simeq N_d\left(\frac{\lambda}{2 \pi}\right)^2
{\frac{\mu^4}{2\sigma}}
\left(1-\frac{\sigma^2}{3N}+\frac{7\sigma^4}{90N^2}
-\frac{31\sigma^6}{1890N^3}+\dots\right)
\end{equation}
for $V_{rad}^{\prime}$. From these expansions we immediately see that with increasing N, as expected,
$V$ and $V_{rad}^{\prime}$ approximate better the potential ${V_{can}}+{\delta}V_{can}$ and the
radiative slope $V_{rad}^{\prime}$ corresponding to a minimal K\"ahler, respectively. Our earlier
discussion concerning the various hybrid inflationary scenarios in minimal or quasi-minimal supergravity
apply to this alternative construction as well, provided $N\gg1$. The cases $N=1, 2$ present, of course,
some novel features \cite{costas} since supergravity corrections to the classical potential appear proportional
to the parameter $\beta$ which essentially controls their strength. Hybrid inflation in these cases is accurately
described by Eqs. (\ref{mass1})-(\ref{mass3}), provided $\sigma_H^2\ll1$.

\setcounter{equation}{0}

\section{``Chaotic" D-term inflation}

Our discussion of the initial condition problem of hybrid inflation led us to conclude that a solution could be
provided by a short Planck-scale inflationary stage which could easily be incorporated into the non-supersymmetric
prototype model. However, from our earlier discussion it should be clear that extending inflationary scenarios in the
context of supergravity is not an easy task especially if the potential during inflation is dominated by the F-term.
Obviously, models with consecutive stages of inflation should be much harder to construct.

As pointed out long time ago the ``obvious" argument concerning the origin of an unacceptably large inflaton mass
in supergravity does not apply if the D-term dominates the potential during inflation \cite{dt}, \cite{st}.
Thus, D-term inflation appears to be an attractive candidate for the early inflationary stage \cite{lt}.
In the previous section we presented realizations of supergravity hybrid inflation in which the contributions of
``decoupled" fields acquiring large vevs through ``anomalous" D-terms play an important role in the suppression
of the inflaton mass. It would be very desirable on the grounds of simplicity and economy that the D-term sector
involving the ``decoupled" fields provides the necessary short inflationary stage in order to resolve the initial
condition problem of hybrid inflation.  

Let us consider again the chiral superfield $Z$ with charge $-1$ under an ``anomalous" $U(1)$ gauge symmetry
and the associated with it D-term
\begin{equation}
V_D=\frac{g^{2}}{2}\left( {K_Z}Z-\xi \right) ^{2}.
\end{equation}
If during some period of time $\left|{K_Z}{Z}\right|\ll\xi$ the D-term potential becomes approximately constant
\begin{equation}
V_D=\frac{1}{2}g^{2}\xi ^{2}
\end{equation}
and on the condition that this constant dominates the potential the universe experiences a period of
quasi-exponential expansion. In the standard D-term inflationary scenario $\left|{K_Z}{Z}\right|$
is kept small because the scalar field $Z$ finds itself lying close to zero trapped in a wrong vacuum.
Such a scenario, in analogy with the hybrid one, requires the presence of at least one additional
field playing the role of the inflaton whose magnitude determines the size and the sign of the mass-squared of $Z$. 
In contrast to the standard scenario the D-term inflationary scenarios that we deal with here involve no other
fields besides $Z$ which necessarily plays the role of the inflaton. Here the term ``chaotic" is simply meant to
indicate that (the canonically normalized field in) $Z$ is not trapped in a wrong vacuum and its initial value does
not have to be small. We are going to consider two such ``chaotic" D-term inflationary scenarios. The one relies
on a choice of the K\"ahler potential of the type encountered in no-scale supergravity \cite{costas}
whereas the other employs more ``conventional" choices of K\"ahler potentials but invokes rather specific values
of the Fayet-Iliopoulos $\xi$ term.

\subsection{Scenarios with K\"ahler potentials of the no-scale type}

Let us assume that the field $Z$ with charge $-1$ under an ``anomalous" $U(1)$ gauge group enters
the K\"ahler potential $K=K_{1}(\left| S\right|^{2})+K_{2}(\left| Z\right| ^{2})$ through a function $K_2$
of the no-scale type
\begin{eqnarray}
   \label{nsc}
K_{2} = {- n}\ln\left(- \ln\left|Z\right|^2\right) 
\qquad
\left( 0 <  \left| Z\right| ^{2} < 1 \right) ,
\end{eqnarray}
where $n$ is an integer. The corresponding K\"ahler manifold is (again) the coset space $SU(1,1)/U(1)$
of constant scalar curvature $2 / n$. Bringing $Z$ to the real axis by a gauge transformation we define
the canonically normalized real scalar field $\zeta$ through the relation
\begin{equation}
e^{\sqrt{\frac{2}{n}}\zeta}\equiv - \frac{n}{\xi  \ln \left|Z\right|^2}.
\end{equation}
Here the integration constant is chosen in a way that simplifies the expression for the ``anomalous"
D-term potential
\begin{equation}
V_{D} = \frac{1}{2}g^2\xi^2\left(e^{\sqrt{\frac{2}{n}}\zeta} -1\right)^2
\end{equation}
which is assumed to be the dominant contribution to the D-term potential. According to our assumption
$Z$ does not enter the superpotential $W$ at all. Moreover, the quantity 
\begin{equation}
\left|K_Z\right|^2\left(K_{ZZ^*}\right)^{-1} =n
\end{equation}
turns out to be field-independent. Thus, the only dependence of the F-term potential $V_F$
on $Z$ is through the exponential factor 
\begin{equation}
e^{K_2} =\left(\frac{\xi}{n}\right)^n e^{\sqrt{2n}\zeta}.
\end{equation}
As a consequence we obtain
\begin{equation}
\frac{\partial V}{\partial \zeta} = -\sqrt{2n}\left(\frac{1}{{\frac{n}{2}}\left
(e^{-\sqrt{\frac{2}{n}}\zeta}-1\right)} -\frac{V_F}{V_D}\right)V_D.
\end{equation}

Let us assume that $\zeta <0$ and that either from the beginning or after some time
\begin{equation}
   \label{ineq}
\frac{V_D}{V_F} \gg \frac{n}{2}\left(e^{-\sqrt{\frac{2}{n}}\zeta}-1\right).
\end{equation}
The above condition is equivalent to
\begin{equation}
\frac{g^2\xi^2}{V_Fe^{-\sqrt{2n}\zeta}} \gg n
\frac{e^{\sqrt{\frac{2}{n}}(n-1)\zeta} - e^{\sqrt{2n}\zeta}}
{\left(1-e^{\sqrt{\frac{2}{n}}\zeta}\right)^2}
\end{equation}
which can be seen to be easily satisfied if $-\zeta \gg 1$ given that $V_Fe^{-\sqrt{2n}\zeta}$ is
$\zeta$-independent. Then, 
\begin{eqnarray}
V\simeq V_D\simeq \frac{g^2\xi^2}{2}, \nonumber \\
-\frac{1}{V}\frac{\partial V}{\partial \zeta} \simeq
2\sqrt{\frac{2}{n}}e^{\sqrt{\frac{2}{n}}\zeta} \ll 1, \nonumber\\
-\frac{1}{V}\frac{\partial^2 V}{\partial \zeta^2} \simeq
{\frac{4}{n}}e^{\sqrt{\frac{2}{n}}\zeta} \ll 1
\end{eqnarray} 
and provided the initial kinetic energy is much smaller than the potential one the universe experiences
a period of quasi-exponential expansion. If during this period $\zeta$ varies in the interval [$\zeta_{in}$,
$\zeta_{f}$] the corresponding number of e-foldings $N(\zeta_{in}, \zeta_{f})$ in the slow-roll approximation is 
\begin{equation}
N(\zeta_{in}, \zeta_{f}) \simeq \frac{n}{4}\left(e^{-\sqrt{\frac{2}{n}}\zeta_{in}}
-e^{-\sqrt{\frac{2}{n}}\zeta_{f}}\right).
\end{equation}
Inflation ends when the slow-roll parameter $\left|\eta\right|$ approaches unity and $\zeta$
acquires the value $\zeta_f$ given by $e^{-\sqrt{\frac{2}{n}}\zeta_f} \simeq \frac{4}{n}$
assuming that Eq.\  (\ref{ineq}) still holds. This is not impossible since the $\zeta$-independent
quantity $V_Fe^{-\sqrt{2n}\zeta}$ depends on other fields whose evolution could easily lead to
a fast decrease of the F-term in the meantime. Finally, $\zeta$ starts performing damped
oscillations around its (only slightly displaced from zero) minimum.

With the choice of $K_{1}(\left| S\right|^{2})$ given by Eq.\  (\ref{KS}) and $K_{2}(\left| Z \right|^{2})$
given by Eq.\  (\ref{nsc}) the parameter $\beta$ becomes
\begin{equation}
  \label{beta2}
\beta= -\frac{2}{N} +n.
\end{equation}
It is remarkable that $\beta$ naturally vanishes along with all supergravity corrections for
$(N, n)=(1, 2)$ or $(N, n)=(2, 1)$. Moreover, if two $Z-$type fields with K\"ahler potentials given
by Eq.\  (\ref{nsc}) and characterized by integers $n_1$ and $n_2$, respectively are employed
\begin{equation}
   \label{beta3}
\beta= -\frac{2}{N} +n_1+n_2.
\end{equation}
Vanishing of $\beta$ is again achieved with the unique choice $N=n_1=n_2=1$. In the last case an early
``chaotic" D-term inflation in two stages is possible which, as we shall see, allows for a completely satisfactory
solution of the initial condition problem of hybrid inflation. A small $\beta$, if desired, can be generated by the
contribution of an additional field with a D-term involving a more ``conventional" K\"ahler potential and a small
$\xi$ parameter \cite{costas}. In such a case supergravity corrections do not vanish but they still remain
suppressed since they are proportional to the small parameter $\beta$.

\subsection{Scenarios realizable for specific values of the $\xi$ term}

Let us consider the double-well potential
\begin{equation}
V=\frac{g^2}{2}\left(\frac{\zeta^2}{2} -\xi\right)^2
\end{equation}
involving the real scalar field $\zeta$ with canonically normalized kinetic term. This is the ``anomalous" D-term
potential of a field $Z$ with a minimal K\"ahler potential which is brought to the real axis ($ReZ=\frac{\zeta}{\sqrt{2}}$)
by a gauge transformation. For $\zeta \gg 1$, as well-known, the equation of motion
\begin{equation}
   \label{em1}
\frac{dv}{d\zeta}+\sqrt{3\left(\frac{1}{2}v^2+V\right)}+\frac{V^{\prime}}{v}=0
\end{equation}
with 
\begin{equation}
v \equiv \dot \zeta \equiv \frac{d\zeta}{dt}
\end{equation}
admits the approximate inflationary slow-roll solution
\begin{equation}
v_{SR}=-\sqrt{\frac{2}{3}}g\zeta.
\end{equation}
It is not difficult to verify that for the specific value
\begin{equation}
\xi=\xi_{c}\equiv \frac{1}{3}
\end{equation}
the energy $E=\frac{1}{2}v^2+V$ calculated for the above solution becomes a ``perfect square"
\begin{equation}
E_{SR}=\frac{1}{2}v_{SR}^2+\frac{g^2}{2}\left(\frac{\zeta^2}{2}-\frac{1}{3}\right)^2
=\frac{g^2}{2}\left(\frac{\zeta^2}{2}+\frac{1}{3}\right)^2
\end{equation}
and the approximate slow-roll solution becomes exact!
Integration of this exact solution 
\begin{equation}
   \label{ex}
v_{e}=-\sqrt{\frac{2}{3}}g\zeta
\end{equation}
leads to
\begin{equation}
\zeta=\zeta_0 e^{-\sqrt{\frac{2}{3}}gt}
\end{equation}
which demonstrates the important fact that $\zeta$ does not oscillate
but vanishes only asymptotically with time.
In the following we refer to $v_e$ as a non-oscillatory solution 
in the sense that it describes a non-oscillatory behavior of $\zeta$.
The number of e-foldings during the variation of $\zeta$ from an initial
value $\zeta_{in}$ to a final value $\zeta_{f}$ is given by the formula
\begin{equation}
N(\zeta_{in}, \zeta_{f})=\int_{t_{in}}^{t_{f}}Hdt=
\frac{1}{8}\left(\zeta_{in}^2-\zeta_{f}^2\right)
+\frac{1}{6} \ln\left|\frac{\zeta_{in}}{\zeta_{f}}\right|.
\end{equation}
Moreover, the condition for inflationary expansion
\begin{equation}
-\frac{\dot H}{H^2}=3\frac{E_k}{E} < 1,
\end{equation}
where $E_k=\frac{1}{2}v^2$ is the kinetic energy and $H=\sqrt{\frac{E}{3}}$,
is violated only for $\frac{\zeta^2}{2}$ in the interval [$r_{-}, r_{+}$]
with $r_{\pm}=\left(1\pm\sqrt{\frac{2}{3}}\right)^2$. 
We see that our special non-oscillatory solution $v_e$ leads to inflation at
both large and small field values with the intervention of only a short
non-inflationary period. For $\left|\zeta\right| \gg 1$ the well-known ``chaotic"
inflationary scenario is realized. For $\left|\zeta\right| \ll 1$  inflation takes
place because the energy density is dominated by the constant
$\frac{1}{2}g^2\xi^2$. If our solution were followed with infinite accuracy
inflation would never end. Fortunately, this is only mathematically possible.

The question that naturally arises is the extent to which an
inflationary scenario relying on the special non-oscillatory
solution $v_e$ is realizable. Stated differently, we should 
investigate whether it is probable at all that the evolution 
of the field $\zeta$ follows, even approximately, Eq.\  (\ref{ex}).
Obviously, this depends on the ``stability" of the non-oscillatory
solution relative to small perturbations. We will call the 
solution $v$  ``stable" if a solution $v+\delta v$ that initially
deviates slightly from it tends to ``approach" $v$ in the sense
that $\left|\frac{\delta v}{v}\right|$ decreases during the evolution.
Keeping only linear terms in $\delta v$ the evolution of an
initial perturbation $\delta v_0$ is described by the relation
\begin{equation}
   \label{pert1}
\left|\frac{\delta v}{v}\right|=\left|\frac{\delta v_0}{v_0}\right|
e^{-\int_{{0}}^{t}{\frac{H}{\dot H}}\frac{d}{dt}\left(\frac{V}{H}\right)dt}
=\left|\frac{\delta v_0}{v_0}\right|
e^{-\int_{{0}}^{t}\left(3-{\frac{\dot H}{H^2}}+\frac{\ddot H}{{\dot H}H}\right)Hdt}.
\end{equation} 
It follows that $\left|\frac{\delta v}{v}\right|$ tends to fall or grow 
depending on whether $\frac{V}{H}$ decreases or increases during
the evolution. It turns out that our exact solution $v_e$ is  ``stable"
for $\frac{\zeta^2}{2} > \frac{1}{3}$ and ``unstable" for
$\frac{\zeta^2}{2} < \frac{1}{3}$. The ``stability" of $v_e$ changes
at  $\frac{\zeta^2}{2} = \frac{1}{3}$ where $V=V^{\prime}=0$.
If the solution $v$ describes slow-roll inflationary expansion,
like in the case of the ``chaotic" inflation at $\left|\zeta\right|\gg 1$,
$\left|\frac{\dot H}{H^2}\right|$,  $\left|\frac{\ddot H}{{\dot H}H}\right|\ll1$
and $\left|\frac{\delta v}{v}\right| \sim e^{-3N}$, where $N$ is the 
number of e-foldings. Consequently, during this period the evolution of
the field $\zeta$ tends to approach the non-oscillatory solution $v_e$.
The size of the deviation from this solution will depend on the duration
of the inflationary period at large $\left|\zeta\right|$. In contrast, during
the inflationary stage at  $\left|\zeta\right| \ll 1$, although $-\frac{\dot H}{H^2} \ll 1$,
the slow-roll parameter $\left|\eta\right| \gg 1$ and
$\frac{\ddot H}{{\dot H}H} \simeq -12$. As a result 
$\left|\frac{\delta v_e}{v_e}\right| \sim e^{9N}$ and any small deviation
from the non-oscillatory solution $v_e$ will start growing.
This will eventually lead to a termination of the asymptotic approach
of $\zeta$ to the origin.  

It is actually possible to make a stronger statement concerning the
importance of our exact non-oscillatory solution. Let us define
the variable $u\equiv \frac{v}{v_e}$, where $v$ is an arbitrary solution
of the equation of motion (\ref{em1})
involving an arbitrary potential $V$ and $v_e$ 
any specific exact solution. It can be shown that the equation of motion
(\ref{em1}) can be rewritten in terms of $u$ as
\begin{equation}
   \label{em2}
u^{\prime}\equiv\frac{du}{d\zeta}= \frac{(u^2-1)}{u}
\left(\frac{V^{\prime}}{v_e^2}\right)
\frac{\sqrt{E}+wu}{\sqrt{E}+\sqrt{E_e}u},
\end{equation}
where
\begin{equation}
w=\sqrt{E_e}+\sqrt{3}\frac{V}{V^{\prime}}v_e=
\sqrt{E_e}\frac{E_e}{V^{\prime}}\left(\frac{V}{E_e}\right)^{\prime},
\end{equation}
\begin{equation}
E=\frac{1}{2}u^2v_e^2+V
\end{equation}
and
\begin{equation}
E_e=\frac{1}{2}v_e^2+V
\end{equation}
is the energy calculated for the specific solution $v_e$. Let us assume that $w\ge 0$.
Then, from the above expression it becomes obvious that for $u>0$ the sign of
$\frac{du}{dV}=\frac{u^{\prime}}{V^{\prime}}$ coincides with the sign of $u-1$.
Stated differently, all solutions $v$ leading to evolution in the same direction as the
solution $v_e$ tend to ``converge" towards $v_e$ or ``diverge" from it depending on
whether the potential energy $V$ decreases or increases during the evolution.
In the case of our exact special solution $v_e$ of Eq.\  (\ref{ex}) $w=\frac{g}{\sqrt{2}}\frac{2}{3}$
is a positive constant. Therefore, the above remarks find immediate application and
our earlier conclusions concerning the ``stability" of $v_e$ are now strengthened
to incorporate larger deviations from it. 

The evolution of the field $\zeta$ in the vicinity of the origin can be reliably investigated
in the approximation $V\simeq \frac{1}{2}g^2\xi^2$ and $V^{\prime}\simeq -g^2\xi\zeta$.  
We see that the slow-roll parameter $\epsilon \equiv \frac{1}{2}\left(\frac{V^{^{\prime }}}
{V}\right) ^{2}\simeq \frac{2}{\xi^2} \zeta^2 \ll 1$ suggesting the existence of solutions
with $E_k \ll V$ on which we concentrate. The slow-roll parameter $\left|\eta\right| \equiv 
\left|\frac{V^{^{\prime \prime }}}{V}\right|\simeq \frac{2}{\xi}$, instead, is not small which
necessitates a more careful treatment of the simplified equation of motion
\begin{equation}
   \label{em3}
\left(\frac{g\xi}{\sqrt{6}}\right)^{-1}\frac{dv}{d\zeta} +3
-\frac{6}{\xi}\left(\frac{g\xi}{\sqrt{6}}\right)\frac{\zeta}{v}=0
\end{equation}
in which the kinetic energy has been neglected relative to the potential one. The above
equation admits the special linear in $\zeta$ solutions
\begin{equation}
v^{\pm} = \rho_{\pm} \left(\frac{g\xi}{\sqrt{6}}\right)\zeta ,
\end{equation}
where 
\begin{equation}
\rho_{\pm} = \frac{3}{2} \left(-1\pm\sqrt{1+\frac{8}{3\xi}}\right).
\end{equation}
The solution $v^{-}$, which coincides for $\xi =\frac{1}{3}$ with the exact special solution $v_e$,
describes an asymptotic approach of $\zeta$ to the origin. The solution $v^{+}$, instead, describes
evolution of $\zeta$ away from the origin. Both are non-slow-roll inflationary solutions with
$H\simeq \frac{g\xi}{\sqrt{6}}$, $-\frac{\dot H}{H^2}\simeq \frac{1}{2}\rho_{\pm}^2\zeta^2 \ll 1$,  
$\frac{\ddot H}{{\dot H}H} \simeq 2\rho_{\pm}$
and 
\begin{equation}
N(\zeta_{in}, \zeta_{f})=\int_{t_{in}}^{t_{f}}Hdt\simeq
\frac{1}{\rho_{\pm}}\ln\left|\frac{\zeta_{f}}{\zeta_{in}}\right|.
\end{equation}
Substituting in Eq.\  (\ref{pert1}) we obtain
\begin{equation}
  \label{pert2}
\left|\frac{\delta v^{\pm}}{v^{\pm}}\right|\simeq
\left|\frac{\delta v_0^{\pm}}{v_0^{\pm}}\right|
e^{\mp(\rho_{+}-\rho_{-})N(\zeta_0,\zeta)}
\simeq \left|\frac{\delta v_0^{\pm}}{v_0^{\pm}}\right|
\left|\frac{\zeta_0}{\zeta}\right|^{\pm\frac{\rho_{+}-\rho_{-}}{\rho_{\pm}} } .
\end{equation} 
We see that $v^{+}$ is ``stable" but $v^{-}$ `` unstable". In particular, for $\xi =\frac{1}{3}$ we have 
$\left|\frac{\delta v^{\pm}}{v^{\pm}}\right|\sim e^{\mp 9N}$ in accordance with our earlier assertion
concerning $v^{-}$.

The simplicity of the equation of motion in the $\left|\zeta\right| \ll 1$ approximation allows us to
assess the importance of the special solutions $v^{\pm}$ with more confidence. Eq.\  (\ref{em3})
can be rewritten as 
\begin{equation}
   \label{em4}
\frac{d\left(\frac{v}{v^{\pm}}\right)^2}{d \ln \zeta^2}
= - \frac{\rho_{\mp}}{\rho_{\pm}}\left(\frac{v}{v^{+}} -1\right)
\left(\frac{v}{v^{-}} -1\right).
\end{equation}
Notice that $ - \frac{\rho_{\mp}}{\rho_{\pm}}>0$.
When $\frac{v}{v^{+}} > 0$, necessarily $\frac{v}{v^{-}} < 0$
and $\zeta^2$ grows with the evolution. Then, $\frac{v}{v^{+}}$
decreases or increases depending on whether
$\frac{v}{v^{+}}$ is larger or smaller than $1$.
In both cases $v$ approaches $v^{+}$.
When $\frac{v}{v^{-}} > 0$, necessarily $\frac{v}{v^{+}} < 0$
and $\zeta^2$ falls with the evolution. Then, $\frac{v}{v^{-}}$
increases or decreases depending on whether
$\frac{v}{v^{-}}$ is larger or smaller than $1$.
As a consequence $v$ departs further from the special solution $v^{-}$. 
In the first case $\zeta$ crosses the origin with non-zero $v$ whereas
in the second case $v$ vanishes at a non-zero value of $\zeta$.
In both cases $\frac{v}{v^{-}}$ flips sign and eventually $v$ tends to $v^{+}$.
Therefore, $v$ approaches $v^{+}$ in all cases but the very special one
that $v=v^{-}$ to begin with. Integration of Eq.\  (\ref{em4}) leads to the general solution
\begin{equation}
\left|v-v^{+}\right|^{\rho_{+}} \left|v-v^{-}\right|^{-\rho_{-}} = A,
\end{equation}
with $A$ a constant, from which we obtain once more Eq.\  (\ref{pert2}) describing
the evolution of small deviations $\delta v^{\pm}$ from the special 
solutions $v^{\pm}$.

In conclusion, starting from a relatively large $\left|\zeta\right|$ the evolution of the field $\zeta$
will approach the solution $v_e$ giving rise to a ``chaotic" inflationary expansion. After a short
break of the inflationary expansion near the minimum of the potential a new inflationary expansion
begins as $\zeta$ approaches the origin. This approach, however, is combined with a gradual
departure from the non-oscillatory solution. Eventually, $\zeta$ will either stop before reaching
the origin or cross the origin with a small speed. Then, a new inflationary expansion begins as
$\zeta$ moves away from the origin following the solution $v^{+}$. The duration of the inflationary
stages at $\left|\zeta\right| \ll 1$ will, of course, depend on the accuracy with which the evolution
of the field $\zeta$ follows the special solution $v_e$ which in turn depends on the duration of the
inflationary stage at $\left|\zeta\right| \gg 1$.  

We should certainly bear in mind, however, that there will inevitably be deviations from the above
mathematical analysis in realistic situations. As reasons for such deviations we could mention
quantum fluctuations, small contributions from other fields which affect the equation of motion
and a small departure of the parameter $\xi$ from the special value $\xi_{c} =\frac{1}{3}$.

The above discussion can be extended to potentials 
\begin{equation}
V=\frac{g^2}{2}\left( {K_Z}Z-\xi \right) ^{2}
\end{equation}
which are ``anomalous" D-terms involving  fields $Z$ with non-minimal K\"ahler potentials.
The cases that we are going to consider here are the K\"ahler manifolds $SU(1,1)/U(1)$ and 
$SU(2)/U(1)$.

Let us first consider the  K\"ahler manifold $SU(1,1)/U(1)$ and make the ``conventional" choice
\begin{equation}
  \label{KZnc}
K(\left| Z\right| ^{2})=-n\ln \left( 1-\frac{\left| Z\right| ^{2}}{n}
\right) \qquad \left( \left| Z\right| ^{2}<n\right) 
\end{equation}
for the K\"ahler potential involving the field $Z$. Here $n$ is an integer. $Z$ can be brought to the real axis
through a gauge transformation and a canonically normalized real scalar field $\zeta$ is defined by the relation
\begin{equation}
ReZ\equiv \sqrt{n} \tanh{\frac{\zeta}{\sqrt{2n}}}.
\end{equation}
Then, the potential becomes
\begin{equation}
V=\frac{g^2}{2}\left(n\sinh^2\frac{\zeta}{\sqrt{2n}}-\xi\right)^2.
\end{equation}
For the specific value 
\begin{eqnarray}
\xi=\xi_{c}\equiv \frac{n}{2}\left(\sqrt{\frac{3n}{3n-4}} -1\right) \qquad (n>1)
\end{eqnarray}
the equation of motion of the field $\zeta$ admits the exact special non-oscillatory solution
\begin{equation}
v_e= - \frac{n}{\sqrt{3n-4}} g \sinh\left({\sqrt{\frac{2}{n}}}{\zeta}\right)
\end{equation}
for which 
\begin{equation}
\int Hdt = - \frac{n}{4}\left\{
\ln \cosh^2\frac{\zeta}{\sqrt{2n}}+
\left(1-\sqrt{1-\frac{4}{3n}}\right) \ln \tanh\frac{\zeta}{\sqrt{2n}}\right\}.
\end{equation}
The solution $v_e$ does not describe inflationary expansion
($-\frac{\dot H}{H^2} \ge 1$) only for $n\sinh^2\frac{\zeta}{\sqrt{2n}}$ in the
interval [$r_{-}$, $r_{+}$] if $n>4$ or in the interval [$r_{-}$, $+\infty$) if
$1<n\le 4$, with $r_{\pm} = \frac{2}{3}\left(1\pm \sqrt{\frac{2}{3}}\right)
\left(\sqrt{1-\frac{4}{3n}} +1\right)^{-1}
\left(\sqrt{1-\frac{4}{3n}}\mp\sqrt{\frac{2}{3}}\right)^{-1}$.
We see that a ``chaotic" inflationary stage at $\left|\zeta\right| \gg 1$
is possible only for $n>4$. This is a power-law inflation with
$H \sim R^{-\frac{4}{n}}$. The solution $v_e$, instead, describes
inflationary expansion for  $\left|\zeta\right| \ll 1$ even when it does not
for $\left|\zeta\right| \gg 1$. The exact solution $v_e$ is ``stable"
or ``unstable" depending on whether $n\sinh^2\frac{\zeta}{\sqrt{2n}}$ 
is larger or smaller than $\xi_{c}$. For $\left|\zeta\right| \gg 1$, 
$-\frac{\dot H}{H^2} \simeq \frac{4}{n}$ and $\frac{\ddot H}{{\dot H}H} \simeq -\frac{8}{n}$,
whereas for $\left|\zeta\right| \ll 1$, $-\frac{\dot H}{H^2} \sim \zeta^2 \ll 1$ and
$\frac{\ddot H}{{\dot H}H} \simeq -6 \left(\sqrt{1-\frac{4}{3n}} +1\right)$.
Moreover, $w=\frac{g}{\sqrt{2}}\frac{2n}{3n-4} > 0$ and from Eq.\  (\ref{em2})
follows that the conclusions concerning the ``stability" of $v_e$ are 
further strengthened. Finally, our analysis of the evolution of $\zeta$
near the origin goes through as well with $\rho_{+}=3\sqrt{1-\frac{4}{3n}}$
and $\rho_{-} = -3\left(\sqrt{1-\frac{4}{3n}}+1\right)$. 
Notice that $\rho_{-} \frac{ \xi_{c}} {\sqrt{6}}= - \sqrt{\frac{2n}{3n-4}}$
and $v_e \to v^{-}$ with $\zeta \to 0$, as expected.

In the case of the $SU(2)/U(1)$ K\"ahler manifold we make the choice 
\begin{equation}
   \label{KZc}
K(\left| Z\right| ^{2})=n\ln \left( 1+\frac{\left| Z\right| ^{2}}{n}
\right)  
\end{equation}
for the K\"ahler potential involving the field $Z$. Here $n$ is an even integer. $Z$ can be
brought to the real axis through a gauge transformation and a canonically normalized real
scalar field $\zeta$ is defined as an angular variable by the relation
\begin{equation}
ReZ\equiv \sqrt{n} \tan{\frac{\zeta}{\sqrt{2n}}}.
\end{equation}
Then, the potential becomes
\begin{equation}
V=\frac{g^2}{2}\left(n\sin^2\frac{\zeta}{\sqrt{2n}}-\xi\right)^2.
\end{equation}
For the specific value 
\begin{eqnarray}
\xi=\xi_{c}\equiv \frac{n}{2}\left(1-\sqrt{\frac{3n}{3n+4}}\right) 
\end{eqnarray}
the equation of motion of the field $\zeta$ admits the exact special non-oscillatory solution
\begin{equation}
v_e= - \frac{n}{\sqrt{3n+4}} g \sin\left({\sqrt{\frac{2}{n}}}{\zeta}\right)
\end{equation}
for which 
\begin{equation}
\int Hdt = - \frac{n}{4}\left\{
- \ln \cos^2\frac{\zeta}{\sqrt{2n}}+
\left(\sqrt{1+\frac{4}{3n}}-1\right)  \ln \left| \tan\frac{\zeta}{\sqrt{2n}}\right|\right\}.
\end{equation}
The solution $v_e$ does not describe inflationary expansion
($-\frac{\dot H}{H^2} \ge 1$) only for $n \sin^2\frac{\zeta}{\sqrt{2n}}$ in the
interval [$r_{-}$, $r_{+}$]  with $r_{\pm} = \frac{2}{3}\left(1\pm \sqrt{\frac{2}{3}}\right)
\left(\sqrt{1+\frac{4}{3n}} +1\right)^{-1}
\left(\sqrt{1+\frac{4}{3n}}\mp\sqrt{\frac{2}{3}}\right)^{-1}$.
It is ``stable" or ``unstable" depending on whether $n \sin^2\frac{\zeta}{\sqrt{2n}}$ 
is larger or smaller than $\xi_{c}$. For $\left|\zeta\right| \simeq \sqrt{\frac{n}{2}}\pi$, 
$-\frac{\dot H}{H^2} \sim \cos^2\frac{\zeta}{\sqrt{2n}} \ll 1$ and 
$\frac{\ddot H}{{\dot H}H} \simeq 6 \left(\sqrt{1+\frac{4}{3n}} -1\right)$,
whereas for $\left|\zeta\right| \ll 1$, $-\frac{\dot H}{H^2} \sim \zeta^2 \ll 1$ and
$\frac{\ddot H}{{\dot H}H} \simeq -6 \left(\sqrt{1+\frac{4}{3n}} +1\right)$.
Moreover, $w=\frac{g}{\sqrt{2}}\frac{2n}{3n+4} > 0$ and from Eq.\  (\ref{em2})
follows that the conclusions concerning the ``stability" of $v_e$ are 
further strengthened. Finally, our analysis of the evolution of $\zeta$
near the origin goes through as well with $\rho_{+}=3\sqrt{1+\frac{4}{3n}}$
and $\rho_{-} = -3\left(\sqrt{1+\frac{4}{3n}}+1\right)$. 
Again, because $\rho_{-} \frac{ \xi_{c}} {\sqrt{6}}= - \sqrt{\frac{2n}{3n+4}}$
we have $v_e \to v^{-}$ as $\zeta \to 0$.

The above discussion concentrates on the ``anomalous" D-term which is assumed to be dominant during
the initial stages of the evolution. This assumption certainly places constraints on the size of the initial
values of all fields present in the model including the $Z$ field itself since they are all involved in the F-term
potential. A noticeable contribution of $Z$ to the F-term potential, although not the only one, is the exponential
factor $e^{K_{2}(\left|Z\right|^2)}$ which becomes $e^{\frac{\zeta^2}{2}}$, $\cosh^{2n}\frac{\zeta}{\sqrt{2n}}$ or
$\cos^{-2n}\frac{\zeta}{\sqrt{2n}}$ depending on whether the K\"ahler potential considered is the minimal one,
the one of Eq.\  (\ref{KZnc})  or of Eq.\  (\ref{KZc}), respectively. As a consequence of the constraints on the initial
value of $\zeta$ only a very short inflation at $\left|\zeta\right| \gg 1$ is allowed which necessitates the additional
short inflationary stage at $\left|\zeta\right| \ll 1$ to complement the required expansion and solve the problem of
the initial conditions.

The presence in the F-term potential of the $Z$ fields involved in the above scenarios affects, as pointed out
earlier, the mass of the inflaton associated with the ``observable" inflation. With the choice of the K\"ahler
potential $K_{1}(\left|S\right|^2)$ for the inflaton field $S$ given by Eq.\  (\ref{KS}) the parameter $\beta$
parametrizing the inflaton mass becomes
\begin{equation}
\beta = -\frac{2}{N} +\Delta\beta .
\end{equation}
Here $\Delta \beta$ is the contribution due to the presence of the $Z$ fields which are assumed to lie in their
true vacua during the ``observable" inflation. For all the cases considered above and assuming that $\xi$
takes the specific values $\xi_{c}$ giving rise to the non-oscillatory solutions $v_e$ we have
\begin{equation}
\Delta \beta = \frac{1}{3} \left(\frac{2}{1+\sqrt{1\pm\frac{4}{3n}}}\right)^2 .
\end{equation}
Here the $+$ sign is chosen if the K\"ahler potential $K_{2}(\left|Z\right|^2)$ is the one of Eq.\  (\ref{KZc})
corresponding to the compact $SU(2)/U(1)$ K\"ahler manifold whereas the $-$ sign is chosen if the K\"ahler
potential $K_{2}(\left|Z\right|^2)$ is the one of Eq.\  (\ref{KZnc}) corresponding to the non-compact
$SU(1,1)/U(1)$ K\"ahler manifold. In the former case $0.254 \lesssim \Delta \beta < \frac{1}{3}$ (for $n\ge 2$)
whereas in the latter $\frac{1}{3} < \Delta \beta \lesssim 0.387$ (for $n\ge 5$). The value $\Delta \beta = \frac{1}{3}$
corresponding to the minimal K\"ahler potential is approached from below or from above as $n \to \infty$.
We see that the choice of the minimal K\"ahler potential for $Z$ combined with the choice $N=6$ for
$K_{1}(\left|S\right|^2)$ leads to a massless inflaton of the ``observable" inflation. Choosing a $SU(1,1)/U(1)$
K\"ahler manifold for $Z$ and $N=6$ for $K_{1}(\left|S\right|^2)$, or a $SU(2)/U(1)$ K\"ahler manifold for $Z$ and
$N=7,8,$ or $9$ for $K_{1}(\left|S\right|^2)$ we obtain numerous models with acceptable values of
$\beta \lesssim 3 \times 10^{-2}$. Notice that in these models, unlike the ones of the previous subsection,
supergravity corrections to the classical potential are not proportional to the small parameter $\beta$.
Consequently, the inflationary potential is steep at large inflaton field values although its slope is smaller
than the slope of $V_{can}+\delta V_{can}$ corresponding to a minimal K\"ahler potential for the inflaton $S$.

\setcounter{equation}{0}

\section{The initial conditions}
In the previous section we pointed out that in certain realizations of supergravity hybrid
inflation involving ``decoupled" fields acquiring large vevs through ``anomalous" D-terms
there is a possibility of an early ``chaotic" D-term inflationary stage which could provide a
solution of the initial condition problem of the ``observable" inflation. The ``chaotic" D-term
inflation was analyzed mathematically but only on the assumption that the effect of the
additional potential terms on the ``anomalous" D-term potential is negligible during the early
stages of the evolution. The extent to which this is actually the case depends crucially on the
form of the part of the K\"ahler potential involving the ``decoupled" fields which play the
dominant role during the ``chaotic" D-term inflation. The scenarios with K\"ahler potentials of
the no-scale type are certainly preferable in this respect and are expected to allow for more
natural initial conditions characterized by larger initial values of the fields involved in the
``observable" hybrid inflationary stage. On the other hand one might argue that K\"ahler potentials
of the no-scale type are perhaps too special and scenarios employing more ``conventional"
choices should also be considered. In any case, the problem of investigating whether a given set
of initial conditions does lead to an acceptable ``observable" inflation following an early stage
of inflationary expansion is too complicated to be addressed intuitively or purely analytically and
a numerical analysis of the field evolution becomes necessary. An additional important factor
complicating our task is the effect of the quantum fluctuations during the early inflationary stage.
These fluctuations place lower bounds on the size of all nearly massless fields and an upper
bound on the energy scale where the early inflation should end (see Eq.\  (\ref{bound1})).

In the following we present the results of a numerical study of the initial conditions which seem
to lead to a successful ``observable" inflation according to various scenarios of supergravity
hybrid inflation. The models are classified in two groups depending on the type of the ``chaotic"
D-term inflation. Throughout the discussion the value $\frac{\Delta T}{T} =6.6 \times 10^{-6}$ is
employed. Moreover, the scalar spectral index is evaluated at the scale corresponding to a
wavenumber $k=0.002$ Mpc$^{-1}$ and denoted $n_{0.002}$.

\subsection{Scenarios with K\"ahler potentials of the no-scale type}

For such scenarios the parameter $\beta$ most naturally vanishes
along with all supergravity corrections to the classical potential
although there is a possibility of generating a small value for $\beta$
through additional $Z-$type fields \cite{costas}. Therefore, we will demonstrate
the solution of the initial condition problem in the context of a scenario
for  ``observable" inflation realizing the so-called supersymmetric 
hybrid inflation in which the slope of the potential is provided entirely
by radiative corrections. We will consider two cases for the value of
the coupling $\lambda$. In the first case $\lambda =4.56 \times 10^{-3}$.
Then, for $N_H\simeq 56$ and $N_d=1$ we obtain $M\simeq 2.258 \times 10^{-3}$
($M\simeq 5.5 \times 10^{15}$ GeV) $\mu \simeq 1.525 \times 10^{-4}$,
$\sigma_H \simeq 6.3 \times 10^{-3}$ and $n_{0.002} \simeq 0.986$,
in good agreement with Eqs.\  (\ref{s4}), (\ref{s2}) and (\ref{s3}). Here inflation ends through the
waterfall mechanism. In the second case $\lambda =0.1$.
Then, for $N_H\simeq 55$ and $N_d=1$ we obtain $M\simeq 2.44 \times 10^{-3}$
($M\simeq 5.94 \times 10^{15}$ GeV) $\mu \simeq 7.71 \times 10^{-4}$,
$\sigma_H \simeq 0.118$ and $n_{0.002} \simeq 0.982$,
in good agreement with Eqs.\  (\ref{s1}), (\ref{s2}) and (\ref{s3}). Here inflation ends through strong
radiative corrections. Notice that for much larger values of the coupling
$\lambda$ the slope $V_{rad}^{\prime}$ of Eq.\  (\ref{sl2}) is no longer approximated
accurately by Eq.\  (\ref{sl1}) and the scenario departs from the globally supersymmetric
radiative one.

We choose to employ two $Z-$type fields $Z_1$ and $Z_2$ with K\"ahler potentials
given by Eq.\  (\ref{nsc}) and characterized by integers $n_1$ and $n_2$ both equal to $1$.
Vanishing of the parameter $\beta$ according to Eq.\  (\ref{beta3}) entails the value $N=1$
for the integer entering the K\"ahler potential of Eq.\  (\ref{KS}) that involves the inflaton field $S$.
The gauge couplings of the ``anomalous" $U(1)$'s are given values $g_1=1$ 
and $g_2=0.3$ with the corresponding $\xi$ parameters taking the values $\xi_1 =1$
and $\xi_2 =\frac{1}{3}$. As a consequence the ``chaotic" D-term inflation takes
place in two stages at energy density values close to $\frac{1}{2} g_1^2\xi_1^2 \simeq 0.5$
and $\frac{1}{2} g_2^2\xi_2^2 \simeq 0.005$, respectively. This way we achieve our goal of an early
inflation starting close to the Planck scale and ending at a sufficiently low energy density satisfying
the bound of Eq.\  (\ref{bound1}). Finally, for the fields $\Phi ,\bar{\Phi}$ we employ the minimal
K\"ahler potential.

In the numerical investigation the initial field values for the scenario with
$\lambda=4.56 \times 10^{-3}$ were chosen to be $\sigma_{0} =3$, 
${{\zeta}_{1}}_{0} ={{\zeta}_{2}}_{0} = -2.6$, $\varphi_{0} = \psi_{0} =3$ 
and for the scenario with $\lambda=0.1$ were chosen to be $\sigma_{0} =2.2$,
${{\zeta}_{1}}_{0} ={{\zeta}_{2}}_{0} = -2.3$, $\varphi_{0} = \psi_{0} =2.2$.
Here $\varphi$ and $\psi$ are canonically normalized real scalar fields defined
by $\Phi=\bar{\Phi}= \frac{1}{2}(\varphi + i \psi)$, assuming that $\Phi ,\bar{\Phi}$ lie 
along such a D-flat direction. In both cases the initial time derivatives of the
fields were assumed to vanish and the initial energy density was $\rho_{0}\simeq 1$.  
In the former case the first stage of inflation starts at $\rho \simeq 1$ and ends at
$\rho \simeq 0.2$ giving approximately $15$ e-foldings whereas the second stage
starts at $\rho\simeq 0.02$ and ends at $\rho \simeq 0.002$ giving approximately
$30$ e-foldings. In the latter case the two inflationary stages take place between
the same energy density values but the number of e-foldings is about $10$ and $23$
during the first and the second stage, respectively. Thus, in our scenario the first
stage is only used to generate natural initial conditions for the second stage and
the second stage to set the initial conditions for the ``observable" inflation.
This becomes possible since the number of e-foldings of the second stage is 
sufficiently large. Notice that the initial field values are considerably larger 
than the size of the quantum fluctuations allowing us to omit a more detailed
discussion of the impact of such effects on the field evolution.  

\subsection{Scenarios realizable for specific values of the $\xi$ term}

From our earlier discussion we know that for such scenarios just one $Z-$type
field is able to give rise to an early inflation which could solve the initial condition
problem of the subsequent ``observable" inflation suppressing at the same time 
the mass of the inflaton associated with the latter. The early inflation takes place
in two stages which are characterized by inflaton field values of size 
$\left|\zeta\right| \gg 1$ the first and $\left|\zeta\right| \ll 1$ the second. In such 
scenarios $\xi$ is of order unity. Thus, the only parameter left, apart from the
initial value $\zeta_{0}$ of the canonically normalized inflaton, to determine 
the energy density range where the early inflation takes place is the coupling
$g$ of the ``anomalous" $U(1)$. Notice that the total number of e-foldings
in such scenarios cannot be large given that $\left|\zeta_{0}\right|$ is severely
constrained by the presence of the additional fields associated with the
``observable" inflation and the requirement that the ``anomalous" D-term dominates
the potential during the early inflation. Moreover, the second stage of the early
inflation contributes a very small number of e-foldings since $\zeta$ cannot
go arbitrarily close to the origin partly because $\left|\zeta_{0}\right|$ cannot be
arbitrarily large and partly because the presence of quantum fluctuations renders
a tiny classical $\zeta$ value during inflation physically meaningless.      
Consequently, unlike the case of K\"ahler potentials of the no-scale type, the first
stage of the early inflation does not merely serve the purpose of generating
natural initial conditions for the second stage and the number of e-foldings
necessary to solve the initial condition problem of ``observable" inflation is the sum
of the e-foldings produced during the first and the second stage of the early
D-term inflation. As a result, the quantum fluctuations generated during the
first stage of the early inflation need to be carefully considered especially
because the initial values of the fields associated with the ``observable" inflation
in such scenarios cannot be large.

In order to keep the size of the quantum fluctuations under control
we choose an initial energy density $\rho_{0}$ somewhat smaller than unity
$\rho_{0} \simeq 0.1$. For the fields
$\Phi \equiv  \frac{1}{2}\left(\varphi_{1} +\psi_{1} + i (\varphi_{2}+\psi_{2})\right), 
\bar\Phi \equiv  \frac{1}{2}\left(\varphi_{1} - \psi_{1} -  i (\varphi_{2}-\psi_{2})\right)$,
where $\varphi_{1}, \varphi_{2}, \psi_{1}, \psi_{2}$ are canonically normalized real
scalar fields, which during the ``observable" inflation are supposed to be very
small we choose initial values $\Phi =0.01(1+i)$, $\bar \Phi = 0$
with $\varphi_{1} = \varphi_{2} = \psi_{1} = \psi_{2} = 0.01$
and vanishing initial time derivatives.
(For $\Phi, \bar\Phi$ we assume a minimal K\"ahler potential.)
These are the fields which are mostly affected by quantum fluctuations during
the early D-term inflation due to their being initially small.
For a field becoming massless at t=0 the distribution of the quasiclassical field
generated during inflation has a variance which can be approximated by
\begin{equation}
\sigma_{fluc}^2(t) \simeq \frac{1}{4 \pi^2} \int_{0}^{t} H^3 dt,
\end{equation}
where H is variable during inflation.
To take into account the effect of the quantum fluctuations generated during the
first stage of the early D-term inflation, which are the most important ones,
we replace in the procedure of numerically solving the equations of motion
the small current values of the fields  $\varphi_{1}, \varphi_{2}, \psi_{1}, \psi_{2}$,
as soon as they become ``massless", with new ones of magnitude $1.65 \sigma_{fluc}$.
These new field values are likely to be larger than the actual ones generated by the quantum
fluctuations with $90$ per cent probability. In the case that the dimensionality $N_d$ of the
representation of the gauge group $G$ to which the fields $\Phi, \bar\Phi$ belong 
is large enough the magnitude of the new field values is chosen to be
just $ \sigma_{fluc}$ which represents the average size of semiclassical
field values generated by quantum fluctuations.
The calculation of the expansion that the scale factor of the universe experiences
is performed numerically from the assumed beginning of the evolution, when
$\rho_{0} \simeq 0.1$, till the onset of the low energy density inflation at 
$\rho \simeq 3\mu^4$. In this calculation we consider two extreme cases concerning
the semiclassical fields generated by quantum fluctuations, namely a case of D-flatness with
$- \varphi_{1} = \varphi_{2} = \psi_{1} = \psi_{2} =a $
(i.e. $ \Phi = ia, \bar \Phi = - a$) 
and a maximally non-D-flat case with
$ \varphi_{1} = \varphi_{2} = \psi_{1} = \psi_{2} =a$
(i.e. $ \Phi = (1+i)a, \bar \Phi =0$).
Here $a$ equals $1.65 \sigma_{fluc}$ or $ \sigma_{fluc}$ depending on whether $N_d=1$
or $N_d \gg 1$. During the early inflation the D-flat case gives a larger number of e-foldings
but also leads to a larger decrease of the size of the inflaton $\sigma$ associated 
with the ``observable" inflation. In both cases we make sure that the expansion is 
sufficient to solve the problem of the initial conditions but also that $\left| \sigma \right|$
remains large enough such that the late inflation is able to begin at $\rho \simeq 3\mu^4$.

A more detailed discussion of the initial conditions depends, of course, on the scenario
chosen for the ``observable" inflation. We have seen that in the models that we consider
the mass-squared of the inflaton $\sigma$ associated with the ``observable" inflation
can be expressed in terms of the parameter 
\begin{equation}
\beta=\frac{m_{\sigma}^2}{\mu^4} =
 - \frac{2}{N} + \frac{1}{3} \left(\frac{2}{1+\sqrt{1\pm\frac{4}{3n}}}\right)^2.
\end{equation}
Thus, the ``observable" inflation scenario in our construction depends on the values
of the integers $N$ and $n$ entering the K\"ahler potential and on whether the K\"ahler 
manifold associated with the field $Z$ is compact (+ sign) or non-compact ($-$ sign).

\begin{table}[t]
\begin{center}
\begin{tabular}{lcccccccc}
\hline\hline
$N_d$ & ~~~$\frac{M}{10^{16} GeV}$~~~ & ~~~$\frac{\mu}{10^{-3}}$~~~& ~~$\lambda$~~  
 & ~~$\sigma_H$~~ &$n_{0.002}$& ~~$g$~~ & ~~$\zeta_{0}$~~ & ~~$\sigma_{0}$~~ \\
\hline
1  & 0.8  & 1.150  &  0.1215 & 0.1631&1.028  &0.043 &4.60 &0.3\\
1  & 1.0  & 1.603  &  0.1524 & 0.2255 &1.070 &0.045 &4.50 &0.3\\
1  & 1.2  & 2.030  &  0.170   & 0.2741 &1.112 &0.045 &4.45&0.4\\
32& 2.86& 2.030  &  0.030   & 0.2743  &1.112 &0.040 &4.75&0.5\\

\hline
\end{tabular}
\end{center}
\caption{Parameters and initial field values of supergravity hybrid inflation scenarios
with a massless inflaton. Everywhere $N=6$ and $n=\infty$.} 
\end{table}

The choice $N=6$ for the K\"ahler potential $K_1(\left|S\right|^2)$ of Eq.\  (\ref{KS}) combined with
a minimal  K\"ahler potential $K_2(\left|Z\right|^2)=\left|Z\right|^2$
(which corresponds to $n= \infty$) gives $\beta=0$ and a ``radiative" scenario which is
well approximated by the ``radiative" solution of Eqs.\  (\ref{rad2})-(\ref{rad4}) 
($f_{1,2}\simeq 1$) with $c=\frac{5}{9}$. 
For such a scenario the scale $M$ is smaller than the MSSM scale
$M\simeq 2.86 \times 10^{16}$ GeV unless $N_d$ is considerably larger than unity.
In Table 1 we present some examples of such ``radiative" scenarios with
$N_d =1$ and one example in which the MSSM scale is obtained for $N_d=32$.
For $N_d=1$ the initial values of the $4$ scalar fields 
$\varphi_{1}, \varphi_{2}, \psi_{1}, \psi_{2}$ are assumed to be equal to $0.01$.
For $N_d=32$, however, we assume that initially the $128$ such fields have 
values equal to $0.003$. As the parameter $\mu$ increases $\sigma_H$, the value 
that the inflaton field $\sigma$ takes at the point where the spectrum
of density perturbations is normalized, increases as well and eventually 
the inflaton field has to be given a larger initial value $\sigma_{0}$.
The spectrum of density perturbations is blue ($n_{0.002} >1$) with the quantity
$\sqrt{N_d}\lambda$ controlling the strength of the radiative corrections
being $\gtrsim 0.1$. However, values of this quantity much larger than the ones of Table 1 would
not be allowed since they would result in an unacceptably large value of the spectral index.

Tables 2 and 3 present several scenarios with the ``observable" inflaton possessing
a non-negligible mass. In the scenarios of Table 2 the K\"ahler manifold associated
with the field $Z$ is the non-compact $SU(1,1)/U(1)$ whereas in the scenarios of 
Table 3 the compact $SU(2)/U(1)$. Throughout, the MSSM scale
$M\simeq 2.86 \times 10^{16}$ GeV is assumed and $N_d=1$.
The initial values of the $4$ fields $\varphi_{1}, \varphi_{2}, \psi_{1}, \psi_{2}$
are taken to be $0.01$ and the initial value of the inflaton $\sigma$ is $0.3$ everywhere.
The spectrum of density perturbations is again blue ($n_{0.002} >1$) but with a rather
weak coupling $\lambda \sim 10^{-2} - 10^{-3}$. In all such scenarios
hybrid inflation ends through the waterfall mechanism.

\begin{table}[t]
\begin{center}
\begin{tabular}{lcccccccc}
\hline\hline
$N$&$n$ & ~~~$\frac{\beta}{10^{-2}}$~~~ & ~~~$\frac{\mu}{10^{-3}}$~~~& ~~$\frac{\lambda}{10^{-2}}$~~  
 & ~~$\sigma_H$~~ &$n_{0.002}$& ~~$g$~~ & ~~$\zeta_{0}$  \\
\hline
6  & 8  & 3.1058 &  1.480  & 1.5916&0.1120&1.081 &0.011 &6.20  \\
6  & 10& 2.4266 &  0.984& 0.7036&0.0675&1.056 &0.012 &6.35\\
6  & 12& 1.9914 &  0.772& 0.4331&0.0515&1.044 &0.013 &6.40  \\
6  & 15& 1.5693 &  0.604& 0.2651&0.0404&1.034 &0.014 &6.40  \\
6  & 23& 1.0027 &  0.409 & 0.1216&0.0292&1.021&0.016 &6.50  \\

\hline
\end{tabular}
\end{center}
\caption{Parameters and initial field values of supergravity hybrid inflation scenarios
with a non-zero inflaton mass. The K\"ahler manifold of the field $Z$ is the non-
compact $SU(1,1)/U(1)$. Everywhere $N_d=1$, $M=2.86\times10^{16}$ GeV and
$\sigma_{0} =0.3$.} 
\end{table}

\begin{table}[t]
\begin{center}
\begin{tabular}{lcccccccc}
\hline\hline
$N$&$n$ & ~~~$\frac{\beta}{10^{-2}}$~~~ & ~~~$\frac{\mu}{10^{-3}}$~~~& ~~$\frac{\lambda}{10^{-2}}$~~  
 & ~~$\sigma_H$~~ &$n_{0.002}$& ~~$g$~~ & ~~$\zeta_{0}$  \\
\hline
9  & 2  & 3.1811&  1.66  & 2.0023&0.1291&1.095 &0.250  &3.03 \\
7  & 14& 3.2632&  1.76 & 2.2508&0.1406&1.098 &0.057  &4.60\\
7  & 12& 3.0294 &  1.42 & 1.4652&0.1058&1.079 &0.060  &4.60 \\
7  & 10& 2.7091 &  1.16 & 0.9778&0.0820&1.066 &0.0635&4.60 \\
7  & 8  & 2.2435 &  0.892& 0.5782&0.0603&1.051&0.070  &4.60 \\
7  & 6  & 1.5040 &  0.581& 0.2453&0.0390&1.033&0.085  &4.50 \\

\hline
\end{tabular}
\end{center}
\caption{Parameters and initial field values of supergravity hybrid inflation scenarios
with a non-zero inflaton mass. The K\"ahler manifold of the field $Z$ is the
compact $SU(2)/U(1)$. Everywhere $N_d=1$, $M=2.86\times10^{16}$ GeV and
$\sigma_{0} =0.3$.} 
\end{table}

In our discussion so far it is understood that $\xi$ takes the specific value
$\xi_{c}$.  Although the existence of the non-oscillatory solution $v_e$
requires a specific value for $\xi$ our mechanism actually exploits
the oscillatory solutions which for some finite period of time lie ``close"
to the non-oscillatory one. Such solutions exhibiting a behavior which 
for some time approximates the exact solution $v_e$ do exist, however, 
even if $\xi$ departs slightly from the specific value $\xi_{c}$. 
We verified that the initial conditions of the scenarios described in Tables 2 and 3 
actually do solve the initial condition problem of the slightly destorted hybrid
inflationary scenarios corresponding to a value of $\xi$ given by 
\begin{equation}
\xi =\xi_{c} \left( 1 + \frac{\delta}{2} \right)
\end{equation}
with $\delta$ belonging to the interval [$ - 10^{-3}, 10^{-3}$]. Thus, a tuning
of a single parameter $\delta$ is able to solve both problems associated with
the ``observable" inflation, namely the problem of the initial conditions
and the problem of suppressing the inflaton mass.       

\setcounter{equation}{0}

\section{Conclusions}

We considered F-term realizations of hybrid inflation in supergravity based on a construction
involving ``decoupled" superheavy fields without superpotential which acquire large vevs
through ``anomalous" D-terms \cite{costas}. This construction allows a suppression of the
inflaton mass but also possesses a built in mechanism giving rise to an early ``chaotic"
D-term inflation which was used to solve the problem of the initial conditions of the ``observable"
inflation. We dealt with two variants of this mechanism. The first employs very special K\"ahler
potentials involving the ``decoupled" fields analogous to the ones encountered in no-scale
supergravity and, as it turns out, is only applicable in connection with very special ``observable"
hybrid inflation scenarios characterized by rather flat potentials in which all supergravity corrections
are proportional to the inflaton mass \cite{costas}. The second,  which is of wider applicability, employs
more ``conventional" K\"ahler potentials involving the ``decoupled" fields but invokes values of the
Fayet-Iliopoulos $\xi$ term which lie in restricted intervals. This second variant of the mechanism was used
in order to solve the initial condition problem of hybrid inflationary scenarios possessing rather steep potentials
which resemble the minimal or quasi-minimal supergravity ones. A numerical investigation of the initial
conditions has been performed using both variants of the ``chaotic" D-term inflation and various
hybrid scenarios for the ``observable" inflation. This investigation revealed that there exist ``reasonable"
initial conditions leading to successful ``observable" hybrid inflation in the context of very specific
supergravity models.
 
\subsection*{}

This research was supported in part by EU under contract
MRTN-CT-2004-503369.

\newpage       
\def\npb#1#2#3{~{\it Nucl. Phys.}~{\bf B~#2},~#3,(#1)}
\def\pl#1#2#3{~{\it Phys. Lett.}~{\bf #2~B},~#3 (#1)}
\def\plb#1#2#3{~{\it Phys. Lett.}~{\bf B~#2},~#3 (#1)}
\def\pr#1#2#3{~{\it Phys. Reports}~{\bf #2},~#3 (#1)}
\def\prd#1#2#3{~{\it Phys. Rev.}~{\bf D~#2},~#3 (#1)}
\def\prl#1#2#3{~{\it Phys. Rev. Lett.}~{\bf #2},~#3 (#1)}
\def\ibid#1#2#3{~{\it ibid.}~{\bf ~#2},~#3 (#1)}
\def\apjl#1#2#3{~{\it Astrophys. J. Lett.}~{\bf #2},~#3 (#1)}


\begin{thebibliography}{}

\bibitem{linde90} A.D. Linde, {\it Particle Physics and Inflationary Cosmology},
Harwood Academic (1990).

\bibitem{cobe} G.F. Smoot {\it et al.}, \apjl{1992}
{396}{L1}; C.L. Bennett {\it et al.}, \apjl{1996}{464}{1}.

\bibitem{hyb} A.D. Linde, \plb{1991}{259}{38}; 
\prd{1994}{49}{748}.

\bibitem{lyth} E.J. Copeland, A.R. Liddle, D.H. Lyth,  
E.D. Stewart and D. Wands,  \prd{1994}{49}{6410}.

\bibitem{in} G. Lazarides, C. Panagiotakopoulos and N.D. Vlachos,
\prd{1996}{54}{1369}; G. Lazarides and N.D. Vlachos, \prd{1997}{56}{4562}.

\bibitem{tet} N. Tetradis,  \prd{1998}{57}{5997}. 

\bibitem{double} C. Panagiotakopoulos and N. Tetradis, 
\prd{1999}{59}{083502}.

\bibitem{pre} K.-I. Izawa, M. Kawasaki and T. Yanagida, \plb{1997}{411}{249}.

\bibitem{dss} G. Dvali, R. Schaefer and Q. Shafi, 
\prl{1994}{73}{1886}.

\bibitem{smooth} G. Lazarides and C. Panagiotakopoulos, \prd{1995}{52}{R559}.

\bibitem{st} E.D. Stewart, \prd{1995}{51}{6847}.

\bibitem{pan} C. Panagiotakopoulos, \prd{1997}{55}{R7335}. 

\bibitem{linde97} A.D. Linde and A. Riotto, \prd{1997}{56}{R1841}. 

\bibitem{pan1} C. Panagiotakopoulos,  \plb{1997}{402}{257}.

\bibitem{lr} D.H. Lyth and A. Riotto,  \pr{1999}{314}{1}.

\bibitem{costas} C. Panagiotakopoulos, \plb{1999}{459}{473}.

\bibitem{dt}  J. A. Casas and C. Munoz, Phys. Lett. B {\bf 216}, 37 (1989);
J. A. Casas, J. M. Moreno, C. Munoz and M. Quiros, Nucl. Phys. B {\bf 328},
272 (1989); P. Binetruy and G. Dvali, Phys. Lett. B {\bf 388}, 241 (1996);
E. Halyo, Phys. Lett. B {\bf 387}, 43 (1996).

\bibitem{lt} G. Lazarides and N. Tetradis, \prd{1998}{58}{123502}.

\end{thebibliography}
\end{document}